\documentclass[12pt]{article}
\usepackage{geometry}
\usepackage{amsmath}
\usepackage{amssymb}
\usepackage{amsmath}
\usepackage{graphicx}
\numberwithin{equation}{section}

\def\magnontext{\mathrel \updownarrow \mkern-11mu \joinrel o}
\def\magnonexp{\mathrel \updownarrow \mkern-6mu \joinrel o}

\def\appendix#1{
  \addtocounter{section}{1}
  \setcounter{equation}{0}
  \renewcommand{\thesection}{\Alph{section}}
 \section*{Appendix \thesection\protect\indent \parbox[t]{11.715cm} {#1}}
  \addcontentsline{toc}{section}{Appendix \thesection\ \ \ #1}
  }

\renewcommand{\thefootnote}{\fnsymbol{footnote}}

\newcommand{\be}{\begin{equation}}
\newcommand{\ee}{\end{equation}}
\newcommand{\ba}{\begin{aligned}}
\newcommand{\ea}{\end{aligned}}

\def\m1{\left(-1\right)^{F_i}}

\makeatletter
\def\sla@#1#2#3#4#5{{%
  \setbox\z@\hbox{$\m@th#4#5$}%
  \setbox\tw@\hbox{$\m@th#4#1$}%
  \dimen4\wd\ifdim\wd\z@<\wd\tw@\tw@\else\z@\fi
  \dimen@\ht\tw@
  \advance\dimen@-\dp\tw@
  \advance\dimen@-\ht\z@
  \advance\dimen@\dp\z@
  \divide\dimen@\tw@
  \advance\dimen@-#3\ht\tw@
  \advance\dimen@-#3\dp\tw@
  \dimen@ii#2\wd\z@  \raise-\dimen@\hbox to\dimen4{%
    \hss\kern\dimen@ii\box\tw@\kern-\dimen@ii\hss}%
  \llap{\hbox to\dimen4{\hss\box\z@\hss}}}}
\def\slashed#1{%
  \expandafter\ifx\csname sla@\string#1\endcsname\relax
    {\mathpalette{\sla@/00}{#1}}%
  \else
    \csname sla@\string#1\endcsname
  \fi}
\makeatother

\newcommand{\beq}{\begin{equation}}
\newcommand{\eeq}{\end{equation}}
\newcommand\beqa{\begin{eqnarray}}
\newcommand\eeqa{\end{eqnarray}}
\newcommand\bea{\begin{array}}
\newcommand\eea{\end{array}}

\newcommand{\nn}{\nonumber}
\newcommand{\neqa}{\nonumber\end{eqnarray}}
\newcommand{\la}{\label}

\newcommand{\color}[1]{}

\def\({\left(}
\def\){\right)}
\def\[{\left[}
\def\]{\right]}

\def\<{\langle}
\def\>{\rangle}

\begin{document}

%%%%%%%%%%%%%%%%%%%%%%%%%%%%%%%%%%%%%
%%%%%%%%%%%%%%%%%%%%%%%%%%%%%%%%%%%%%
%%%%%%%%%%%%%%%%%%%%%%%%%%%%%%%%%%%%%

\thispagestyle{empty}
\begin{flushright}\footnotesize
%\texttt{arxiv:yymm.nnnn}\\
%\texttt{CALT-68-NNNN}\\
\texttt{LPTENS 08/30}\\
%\texttt{SPhT-t08/NNN}\\
\texttt{NSF-KITP-08-93}\\
\vspace{2.1cm}
\end{flushright}

\renewcommand{\thefootnote}{\fnsymbol{footnote}}
\setcounter{footnote}{0}

\begin{center}
{\Large\textbf{\mathversion{bold} Toy models for wrapping effects}\par}

\vspace{2.1cm}

\textrm{Jo\~ao Penedones$^{\alpha}$ and Pedro Vieira$^{\beta}$}
\vspace{1.2cm}

\textit{$^{\alpha}$ 
Kavli Institute for Theoretical Physics\\
University of California, Santa Barbara, CA 93106-4030, USA} \\
\texttt{penedon@kitp.ucsb.edu}
 \vspace{3mm}

\textit{$^{\beta}$ Laboratoire de Physique Th\'eorique
de l'Ecole Normale Sup\'erieure et l'Universit\'e Paris-VI, Paris,
75231, France;  Departamento de F\'\i sica e Centro de F\'\i sica do
Porto Faculdade de Ci\^encias da Universidade do Porto Rua do Campo
Alegre, 687, \,4169-007 Porto, Portugal} \\
\texttt{pedrogvieira@gmail.com}
\vspace{3mm}

%%%%%%%%

\par\vspace{1cm}

\textbf{Abstract}\vspace{5mm}
\end{center}

\noindent
The anomalous dimensions of local single trace gauge invariant operators in $\mathcal{N}=4$ supersymmetric Yang-Mills theory can be computed by diagonalizing a long range integrable Hamiltonian by means of a perturbative asymptotic Bethe ansatz. This formalism breaks down when the number of fields of the composite operator is smaller than the range of the Hamiltonian which coincides with  
the order in perturbation theory at study.  We analyze two spin chain toy models which might shed some light on the physics behind these wrapping effects. One of them, the Hubbard model, is known to be closely related to $\mathcal{N}=4$ SYM. 
In this example, we find that the knowledge of the effective spin chain description is insufficient to reconstruct the finite size effects of the underlying electron theory. We compute the wrapping corrections for generic states and relate them to a Luscher like approach. The second toy models are long range integrable Hamiltonians built from the standard algebraic Bethe anstatz formalism. This construction is valid for any symmetry group. In particular, for non-compact groups it exhibits an interesting relation between wrapping interactions and transcendentality.

\vspace*{\fill}

\setcounter{page}{1}
\renewcommand{\thefootnote}{\arabic{footnote}}
\setcounter{footnote}{0}

\newpage

%%%%%%%%%%%%%%%%%%%%%%%%%%%%%%%%%%%%%
%%%%%%%%%%%%%%%%%%%%%%%%%%%%%%%%%%%%%
%%%%%%%%%%%%%%%%%%%%%%%%%%%%%%%%%%%%%

\tableofcontents

\newpage

%%%%%%%%%%%%%%%%%%%%%%%%%%%%%%%%%%%%%
%%%%%%%%%%%%%%%%%%%%%%%%%%%%%%%%%%%%%
%%%%%%%%%%%%%%%%%%%%%%%%%%%%%%%%%%%%%
\section{Introduction and discussion}

\begin{figure} [t]   \centering
        \resizebox{120mm}{!}{\includegraphics{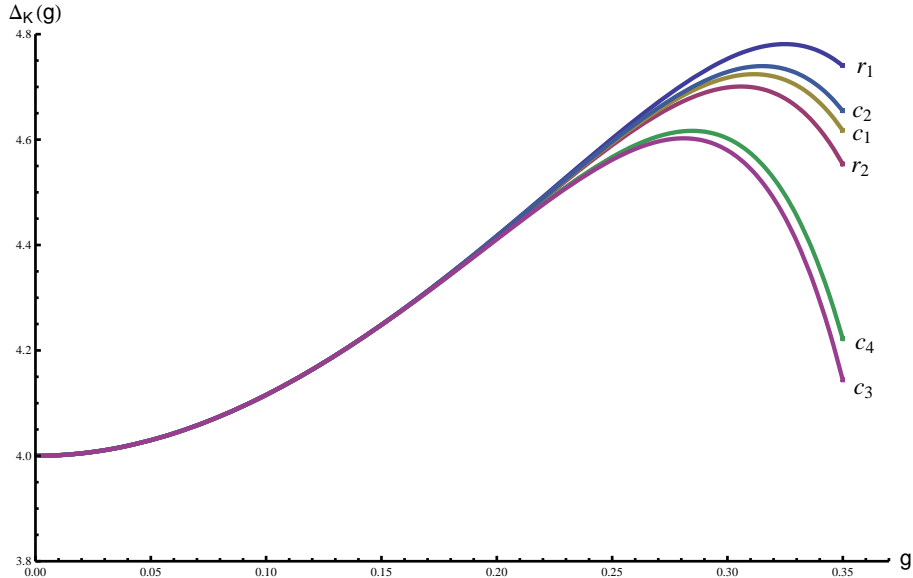}}
\caption{ \footnotesize  Results and conjectures for the scaling dimension $\Delta(g)=4+12g^2-48g^4+336g^6-a g^8$, of the Konishi operator up to four loops, when wrapping interactions first appear. The lines $r_1$ and $r_2$ correspond to the two recent (disagreeing) computations in \cite{w1,w2} for which $a=2607+28 \zeta(3)+140 \zeta(5)$ and $-2584+384 \zeta(3)-1440 \zeta(5)$ respectively. The first one was done using superspace techniques whereas the latter used component formalism making the comparison between these two laborious computations far from easy. 
These computations differ from all previous conjectures by the presence of $\zeta(5)$. $c_1$ is the most recent conjecture \cite{Kotikov:2007cy}, which is based on some transcendentality observations and pomeron considerations and predicts $a=-5307/2+564 \zeta(3)$. Conjectures $c_{2}$ and $c_{3}$ appear in \cite{Beisert:2006ez}. The former corresponds to $a=5640 + 288 \zeta(3)$ and  would be the value of the anomalous dimension of the Konishi operator if we were to believe the Bethe ansatz equation beyond their natural limit of validity and is therefore a very unlikely proposal \cite{Kotikov:2007cy}. $c_3$ is the anomalous dimension whose transcendental part is that given by the BAE while the rational part is that predicted by the Hubbard model and has therefore $a=5088+288 \zeta(3)$. Finally $c_4$ with $a=5088$ would be the result predicted by the Hubbard model \cite{Hubbard} which we now know only reproduces the good BAE up to $3$ loops. 
}\label{fig1}
\end{figure}

In \cite{MZ} the one-loop spectrum of single trace local gauge invariant operators made out of the scalars of $\mathcal{N}=4$ supersymmetric Yang-Mills theory was reduced to that of a nearest neighbors integrable Hamiltonian with $SO(6)$ symmetry. In particular, single trace operators made out of two complex scalars $X$ and $Z$ were mapped to states in a one dimensional spin $1/2$ ring, 
\beq
{\rm tr} \(ZZ\dots Z X Z  \dots Z X Z \dots ZZ\)\,\,\, \longleftrightarrow  \,\,\, | \uparrow \uparrow \dots \uparrow \downarrow \uparrow \dots  \uparrow \downarrow \uparrow \dots  \uparrow \uparrow\, \rangle  \la{example} \
\eeq
Soon after it was understood that integrability persists for the full set of $PSU(2,2|4)$ fields \cite{Beisert:2003jj,Beisert:2003yb} and at higher orders in perturbation theory \cite{Beisert:2004hm} where the Hamiltonian becomes long ranged with the range being the order in perturbation theory one considers. Later on, inspired by string theory data \cite{Kazakov:2004qf,Beisert:2005bm,Arutyunov:2004vx}, the full $PSU(2,2|4)$ Bethe equations were proposed \cite{BS} and the solutions to these equations are believed to yield the spectrum of generic length $L$ operators up to order $g^{2L}$. At this order the interactions wrap the single trace operator and invalidate the use of the Bethe ansatz formalism. To achieve such remarkable point where the spectrum of long operators is believed to be known, a crucial step was required. Namely the idea of looking at operators like (\ref{example})  as a vacuum (the $Z$ fields) on top of which particles (in this example the $X$ fields) propagate \cite{Staudacher:2004tk}. In this language the relevant object becomes the $S$-matrix scattering these particles, also known as magnons. This $S$-matrix is  $SU(2|2)^2$ extended symmetric \cite{Beisert:2005tm} and it turns out that symmetry alone almost fixes (up to an overall function) the $\(4^4\)^2$ entries of this  matrix \cite{Beisert:2005tm,Beisert:2006qh,Arutyunov:2006yd}. The unfixed overall scalar factor has also been conjectured in \cite{Beisert:2006ez,Beisert:2006ib}.  Knowing the $S$-matrix of the theory it is then possible to write down the Bethe equations quantizing their momenta and, knowing the respective dispersion relation, to compute their energy. For example, states made out of two magnons will be given by
\beq
\sum_{n_1 \ll n_2} \(e^{ip_1 n_1 +ip_2 n_2} +S(p_1,p_2) e^{ip_1 n_2 +ip_2 n_1} \)| n_1,n_2 \rangle +\dots \dots \la{magnons}
\eeq 
where $| n_1,n_2 \rangle$ represents the state with $X$ fields in the $n_1$'th and $n_2$'th positions in a sea of $L-2$ $Z$ fields. The dots correspond to a non-trivial part of the eigenstate in the boundary of the asymptotic region when $n_1$ is not very far from $n_2$ and the magnons are strongly interacting.  The momenta are then quantized via the Bethe equations 
\beq
e^{ip_1 L}=S(p_1,p_2) \,\,\, , \,\,\, e^{ip_2 L}=S(p_2,p_1) \nn
\eeq
which physically simply mean that the phase acquired by each magnon when going around the ring equals the free propagation phase $p L$ plus the phase shift due to scattering with the other magnons. The spectrum is then given by the sum of energies of the individual magnons as $\Delta-L+2=\epsilon_\infty(p_1)+\epsilon_\infty(p_2)+\mathcal{O}(g^{2L})$ where the infinite volume dispersion relation, also fixed by symmetry \cite{Beisert:2005tm,Arutyunov:2006ak,Arutyunov:2006yd}, is given by
\beq
\epsilon_\infty(p)=\sqrt{1+16 g^2 \sin^2\frac{p}{2} } \nn \,.
\eeq
The simplest possible $2$ magnon state is the well known Konishi operator 
\beq
|K\rangle = | \downarrow \uparrow \downarrow \uparrow\, \rangle - | \downarrow  \downarrow \uparrow \uparrow \,\rangle \la{Konishi}
\eeq
whose dimension can be computed from the known Bethe ansatz equations \cite{Beisert:2004hm,BS} up to order $g^{2L}=g^8$, 
\beq
\Delta_{K} =4+12g^2-48g^4+336g^6+\mathcal{O}(g^8) \,. \nn
\eeq
At order $g^8$ wrapping interactions appear and the techniques at hand do not suffice to tackle this computation. Still there are already two possible results in the literature \cite{w1,w2} (see also \cite{Eden:2007rd}) where the $g^8$ coefficient was computed by direct evaluation of field theory Feynman diagrams. In figure \ref{fig1} we plot the several computations, conjectures and speculative guesses, for the Konishi anomalous dimension up to four loops.

Optimistically  one might expect to find some extra integrable structure in $\mathcal{N}=4$ SYM which would allow one to treat the gauge invariant states beyond the perturbative asymptotic Bethe ansatz regime.
A particularly appealing possibility would be that some extra hidden local degrees of freedom exist and the long range interactions we perceive would rather be the effect of integrating out these fundamental degrees of freedom. This scenario finds compelling evidence at strong coupling in \cite{Mann:2005ab,Gromov:2006dh,Gromov:2006cq,Gromov:2007cd}, at weak coupling in \cite{Hubbard} and for general coupling in \cite{Rej:2007vm}. 

In \cite{Mann:2005ab,Gromov:2006dh,Gromov:2006cq,Gromov:2007cd} quantum sigma models describing the $S^n$ subsector of $AdS_5\times S^5$ type IIB superstring were seen to reproduce the long range conjectured AFS string Bethe equations \cite{Arutyunov:2004vx} at strong coupling when the rapidities ($\theta$'s) of the relativistic particles were integrated out thus leaving an effective Hamiltonian for the isospin degrees of freedom.

In \cite{Hubbard} the BDS equations \cite{Beisert:2004hm}, which are known to describe the $SU(2)$ sector of the supersymmetric gauge theory spectrum up to three loops, were shown to be equivalent to the Hubbard model at half filling where again integrating out the momenta ($q$'s) of the electrons yields an effective long range Hamiltonian with $SU(2)$ symmetry for the spins of the electrons. 

In the Hubbard model the effective magnons appearing in  (\ref{magnons}) can be understood as bound states of empty sites ($o$) and doubly occupied sites ($\updownarrow$). As we will describe below, if, in the spirit of \cite{Ambjorn:2005wa,Janik:2007wt,Hatsuda:2008gd,Gromov:2008ie,Heller:2008at}, we want to diagrammatically compute the finite size corrections to the effective magnon theory coming from the Hubbard model using the Luscher approach \cite{Luscher:1985dn,Klassen:1990ub}, then we need to take into account the fact that the magnon is a bound state of two fundamental particles rather than a fundamental excitation itself. For example, as discussed in section \ref{Luscher} the leading finite size corrections to the magnon dispersion relation can be reproduced for \textit{any} value of the coupling $g$ from the expression
\beq
\delta \epsilon(p)=\sum_{\sigma=o,\updownarrow} \( \frac{1}{2}\, {\rm Res}_{q} \,e^{i L(q-\phi_\sigma)}
 \(\varepsilon'(q)-\epsilon_\infty'(p)\) S_{\magnonexp,\sigma}^{\magnonexp,\sigma}(p,q) +c.c. \)\la{deltaepsilonintro}
\eeq
where, in order to reproduce the correct result, we must use the scattering matrix between a magnon bound state $\magnontext$ and its fundamental constituents $o$ and $\updownarrow$.

Curiously, Janik and Lukowski \cite{Janik:2007wt} computed the leading finite size correction to the Hubbard magnon energy at large $g=-t/U$, using instead the magnon-magnon scattering matrix and still got a sensible result (correct up to a factor of $2$ -- see equation (71) in \cite{Janik:2007wt}). Physically this makes sense because at strong coupling $g$ -- which from the Hubbard point of view corresponds to weak interaction strength $U$ compared to the electron hopping kinetic energy $t$ -- the magnons are a weakly bound pair of $o$ and $\updownarrow$ and when we scatter the magnon against another magnon we are effectively scattering it against two fundamental particles! However, as the coupling decreases, the magnon-magnon Luscher formula with BDS magnon-magnon S-matrix will no longer provide the correct Hubbard result \footnote{Let us also remark that considering also the contribution from the S-matrix between magnon and bound state of magnons (which exist in the BDS theory) in the Luscher formalism does not cure this problem. At most, these contributions can reproduce the higher winding number diagrams of the fundamental constituents $o$ and $\updownarrow$ only in the strong coupling regime $g\gg 1$.
The reason is again that in this regime a bound state of $b$ magnons is almost like $2 b$ free fundamental particles. }. As explained in section \ref{Luscher} the two results will agree to leading order -- when the correction is of order $1/g$ -- and start disagreeing at next to subleading order -- at order $1/g^3$. 
Could we be in a similar situation in $\mathcal{N}=4$ SYM? There, the finite size corrections \cite{Arutyunov:2006gs,Astolfi:2007uz,Minahan:2008re} to the Giant Magnon \cite{Hofman:2006xt} were reproduced at leading order \cite{Janik:2007wt,Hatsuda:2008gd} and at next to leading order \cite{Gromov:2008ie,Heller:2008at} but no two loop computation is available. Bearing in mind what happens in the Hubbard model it is not completely inconceivable that at this loop order the Luscher results based on the lightcone S-matrix \cite{Beisert:2005tm,Arutyunov:2006yd} start failing. On the other hand from the string worldsheet point of view this scenario would certainly be intriguing. 

In the opposite regime, at weak coupling $g$, expression (\ref{deltaepsilonintro}) gives precisely the correct result whereas the use of the magnon-magnon BDS S-matrix is completely inappropriate because in this regime the elementary particles that make the magnon are highly bound.  If this qualitative  structure persists in the full $\mathcal{N}=4$ theory then it explains why a naive computation of the Luscher terms at weak coupling seems to never yield any sort of transcendental numbers such as $\zeta(3)$ or $\zeta(5)$ which typically appear in the computations of \cite{w1,w2}. The reason would be that in order  to probe the weak coupling limit of the theory the knowledge of the magnon constituents would be essential \footnote{On the other hand, if the magnons in the light-cone gauged string theory can be thought of as fundamental particles -- contrary to what happens in the Hubbard model -- then a priori we should indeed sum over all these infinite number of bound states and of course infinite sums can eventually lead to transcendental numbers. This seems the only way out to find the good transcendental Konishi anomalous dimension from a weak coupling computation based on the Luscher formulae. We acknowledge N.~Gromov, V.~Kazakov and K.~Zarembo for discussions on these issues.}. 

In particular it seems clear that a thermodynamic Bethe ansatz (TBA) approach to the BDS equations would \textit{not} recover the Hubbard finite size corrections (because in particular the TBA approach always reduces to the Luscher formulae at large radius and as we explained those only work when we take into account the scattering of the magnons with the fundamental Hubbard electrons). In the context of the full $\mathcal{N}=4$, the TBA program is being carried out in \cite{TBA1}, \cite{TBA2} still with inconclusive results.

Still in the context of the Hubbard model we analyzed in sections \ref{corrected} and \ref{genluscher} how wrapping interactions manifest themselves for many particle states. In the $\mathcal{N}=4$ context this is an unavoidable question if we want to understand the full anomalous dimensions of small operators such as the Konishi operator (\ref{Konishi}). There are two different kind of effects one needs to take into account to study wrapping corrections to many particle states. On the one hand the energy of the state as a function of the magnon momenta changes when the theory is put in finite volume and this leads to a Luscher type correction which for the Hubbard magnons reads
\beqa
\delta E_{Luscher} =\frac{1}{2}\sum_{n=1}^M \sum_{\sigma=o,\updownarrow} \int_{\mathcal{C}_n} \frac{dq}{2\pi i} \(\varepsilon'(q)-\epsilon_\infty'(p_n)\)\,e^{i \(q -\phi_o \)L}\prod_{m=1}^{M} 
S_{\magnonexp,\sigma}^{\magnonexp,\sigma}(p_m,q) + c.c.  \nn\,,
\eeqa 
generalizing (\ref{deltaepsilonintro}). Led by the striking simplicity of this expression we conjecture a generalization of the Luscher formula for many particle states in integrable two dimensional models in (\ref{conjecture}).

The second effect we need to take into account is the fact that due to the wrapping interactions the quantization conditions -- that is the Bethe ansatz equations -- for the magnon momenta receive corrections and thus the momenta are slightly shifted when wrapping interactions are taken into account. For example, BDS equations \cite{Beisert:2004hm} can be \textit{dressed} to 
\beqa
\(\frac{x_n^+}{x_n^-}\)^L = \prod_{m\neq n} \frac{u_n-u_m+i}{u_n-u_m-i} \,e^{i \phi_{nm}} \nn
\eeqa
in such a way that the leading wrapping interactions are taken into account. Here $\phi_{nm}$ is a \textit{wrapping dressing kernel} described in section \ref{corrected}.

As explained in section \ref{BDSper}, a particularly curious feature of the computations of the weak coupling finite size corrections  is that the leading wrapping correction, of order $g^{2L}$, to a state whose magnons' momenta are $p_j\simeq p_j^{(0)}+g^2 p_j^{(1)}+\dots+g^{2L} p_j^{(L)}$ only depends on the leading values $p_j^{(0)}$.
This is natural from the point of view of the Luscher computations which are basically a smart way to organize the two dimensional perturbative expansion using the two dimensional $S$-matrix. In this formalism the exponentials which appear in the several integrands are automatically of order $g^{2L}$ and thus the rest of the integrand, including the S-matrix and the dispersion relations, can be treated at $g^0$ order. 
 In particular we can easily compute the wrapping correction to any many magnon BDS state without performing any iteration of the BDS equations up to order $g^{2L}$.  
In  $\mathcal{N}=4$, this should be related to the fact that to compute the wrapping corrections to the Konishi operator we isolate the appropriate wrapping diagrams and only keep some information about the lower order graphs \cite{w1,w2}. If the four dimensional wrapping Feynman diagrams in \cite{w1,w2,Eden:2007rd} could be re-written in a two dimensional language this could be the key to understanding the structure of an hypothetical hidden level of fundamental particles.

In section \ref{Toy2} we consider a completely different type of (toy) model where wrapping interactions are under control. Namely we study long ranged Hamiltonians coming from a transfer matrix algebraic construction \textit{\`a la Leningrad school}. In the algebraic Bethe ansatz formalism the fundamental object is the transfer matrix, which is a trace of product of $R$-matrices, 
\beq
T(u)={\rm tr}_{aux} \,R_L(u) \otimes \dots   \otimes R_{1}(u) \la{Tintro}\,,
\eeq
where the $R$-matrices are (simple) matrices obeying the Yang-Baxter relation and acting on the product of an auxiliary space (common for all $R$-matrices in this expression) and a physical space $h_n$. The full Hilbert space of a $L$-site spin chain is given by the tensor product 
\beq
\mathcal{H}=h_1 \otimes \dots \otimes h_L \,. \la{Hilbert}
\eeq
The transfer matrix is then an operator acting on the full Hilbert space (and by definition scalar w.r.t. the auxiliary space). The algebraic Bethe ansatz program yields us the spectrum of the transfer matrix $T(u)$. The diagonalization of this object is of great importance because, as we review in section \ref{Toy2}, by taking $n$ derivatives of the logarithm of the transfer matrix $T$ at $u=0$ we can generate integrable Hamiltonians of range $n$. In particular if we take more than $L$ derivatives of this object we will start generating Hamiltonians which are long ranged, contain wrapping interactions and still, \textit{by construction}, are integrable and diagonalized through a set of Bethe equations. For example if we consider
\beq
\hat H= \left.\frac{1}{4i}\sum_{n=1}^{\infty} \, a_n\,\frac{g^{2n} }{n!} \frac{d^n}{d\lambda^n} \log \hat T(\lambda) \right|_{\lambda=0} +h.c.  \la{Hgenintro}
\eeq
then we will have an Hamiltonian which at order $g^{2n}$ is of range $n$ and whose spectrum is given by a sum of individual dispersion relations plus a wrapping correction which starts precisely at order $g^{2L}$,
\beq
\hat H=\sum_{j=1}^M\epsilon(p_j) + E_{wrapping}(p_1,\dots,p_M) \nn
\eeq
This behavior is probably completely generic and we considered Hamiltonians of the form (\ref{Hgenintro}) with $SU(2)$, $SU(N)$ and $SL(2)$ symmetry. 

In the $SL(2)$ case we found the following curious behavior:
suppose we consider a Hamiltonian of type (\ref{Hgenintro}) with $a_n$ some algebraic numbers. Then the dispersion relation truncated at a given order $g^{2n}$ is a rational function of these algebraic numbers and of the Bethe roots $u_j=\frac{1}{2}\cot\frac{p_j}{2}$, which are quantized via a set of polynomial equations. Thus the contribution to the spectrum of the $\sum_{j=1}^M\epsilon(p_j) $ term will be given by some algebraic quantity. On the other hand, precisely at order $g^{2L}$ the wrapping corrections enter the game and those are given by an infinite sum (\ref{wrapsl2}) of algebraic functions of the Bethe roots $u_j$. Typically they will give rise to transcendental contributions! 

As an example, in the table below we listed a couple of energies of some two magnon states up to order $g^{2L}$ and for the simplest choice of Hamiltonian with $a_n=1$ we get 
\beq
\begin{array}{c|c|c|c}
L &u_{1,2} & \epsilon(p_1)+\epsilon(p_2) & E_{wrapping}(p_1,p_2) \\ \hline 
3 &  \frac{\sqrt{3}}{5}\pm \frac{\sqrt{7}}{10} & \frac{11 g^2}{8}+\frac{13 \sqrt{3}
   g^4}{32}-\frac{5
   g^6}{6}  & \frac{1}{32} \left(-26+3 \pi ^2-4
   \zeta (3)\right) g^6\\
  4 & \frac{1}{3}\pm \frac{\sqrt{7}}{6} & \frac{5
   g^2}{4}-\frac{g^4}{8}-\frac{19
   g^6}{24}+\frac{3
   g^8}{2} & \(\frac{41}{32}-\frac{5 \pi
   ^2}{48}-\frac{\pi ^4}{360}\) g^8 \\
   6 & \pm \frac{1}{2}+\frac{\sqrt{2}}{2} &g^2+\frac{g^4}{\sqrt{2}}-\frac{g^6}{
   3}-\sqrt{2} g^8-\frac{4
   g^{10}}{5}+\frac{5 \sqrt{2}
   g^{12}}{3} &\frac{-1}{2\sqrt{2}}\(7-4 \zeta (3)-2 \zeta (5)\)   g^{12}
\end{array} \la{table}
\eeq
It would be very interesting to explore this connection between  transfer matrices of non-compact groups and transcendentality in the context of $\mathcal{N}=4$ supersymmetric Yang-Mills theory. Perhaps this could provide us with important hints about the origin of the dressing factor which is populated by transcendental numbers. 

This paper is organized as follows: After introducing the Hubbard model in section \ref{hub} and reviewing the magnon description in section \ref{magnon} we re-derive the exact finite volume dispersion relation by means of Feynman diagrams in section \ref{diag}. In section \ref{Luscher} we explain how the leading finite size corrections can be understood from a Luscher type approach. In section \ref{generalwrap} we study wrapping effects for general many particle states. In particular we review how the BDS equations follow from the Lieb-Wu equations, explain how then can be upgraded to include the leading wrapping corrections (section \ref{corrected}) and analyze the analogue of the Luscher formulas for many magnon states (section \ref{genluscher}). Section \ref{Toy2} is devoted to the study of integrable long ranged Hamiltonians derived from an algebraic Bethe ansatz formalism and in section \ref{Toy2gen} we explore some generalizations of this construction and present a curious non-compact long ranged Hamiltonian where transcendentality and wrapping are intimately related.

\section{The Hubbard Model} \la{hub}
The one dimensional Hubbard model describes spin $1/2$ electrons moving in a periodic lattice with $L$ sites.  The electrons can hop between neighboring sites and there is a repulsive (or attractive depending on the sign) potential when two electrons (with opposite spin) occupy the same lattice site. Obviously, due to Pauli exclusion principle no two equal spin electrons can ever occupy the same position. At half filling, when the number of electrons equals the number of sites, each electron will tend to occupy a site in the lattice due to the repulsive potential. We can then study an effective Hamiltonian for the spins alone \cite{alemaes}. It will be a long ranged Hamiltonian where
the interactions correspond to \textit{virtual} processes where electrons hop there and back eventually changing spin in the process. In \cite{Hubbard} this effective Hamiltonian was identified with the long range Hamiltonian of $\mathcal{N}=4$ supersymmetric Yang-Mills theory. This identification is correct up to three loops but fails beyond that. Still, this is an instructive toy model since wrapping interactions, due to electrons making loops around the ring, are perfectly under. In this section we will study them, give them a diagrammatic description and understand how they fit into the usual field theoretical Luscher treatment of finite size corrections.  We will also understand how to modify the effective Bethe equations for the magnons of the effective spin theory  so that they reproduce the (leading) wrapping effects.

A quite useful alternative description of the relevant degrees of freedom in the Hubbard model is obtained by performing a so called Shiba duality. It amounts to thinking of the Hilbert space as that where $N-M$ vacancies $o$ and $M$ double occupancies $\updownarrow$ move in a ferromagnetic vacuum with $L$ up spins\footnote{The number of vacancies $o$ and double occupancies $\updownarrow$ is related to the number of up and down spins as $N_{\updownarrow}=N_\downarrow$, $L-N_o=N_\uparrow$. }. In this description the Hamiltonian reads 
\beqa
H=-t\sum_{i=1}^{L} \sum_{\sigma=o,\updownarrow} \(e^{i\phi_\sigma} a_{i,\sigma}^\dagger a_{i+1,\sigma}  +h.c.\)
-U \sum_{i=1}^L a^\dagger_{i,o}a_{i,o}a^\dagger_{i,\updownarrow}a_{i,\updownarrow} \la{HHub}
\eeqa
where, following \cite{Hubbard}, we have introduced some extra twists $\phi_\sigma$ in the Hamiltonian which can be thought of as a sort of magnetic flux  inducing additional phases in the electron wave function as it moves around the chain. As explained in \cite{Hubbard} and reviewed below these twists can be used to delay the wrapping corrections to the effective spin theory.

The complete spectrum of this Hamiltonian can be obtained as
\beq
E=\sum_{n=1}^{N} \varepsilon(q_n)\,\,\, , \,\,\, \varepsilon(q)\equiv -2t \cos(q) \la{E1}
\eeq
where the momenta are quantized through the solution of the Lieb-Wu \cite{LiebWu} equations,
\beqa
e^{i\(q_n-\phi_o \)L}&=&
\prod_{j=1}^{M}\frac{u_j-2 g \sin(q_n)-i/2}{u_j-2 g \sin(q_n)+i/2} \,\,\,\, , \,\,\,\, n=1,\dots,N \,,\\
e^{i (\phi_o-\phi_\updownarrow)L }\prod_{n=1}^{N} \frac{u_j-2 g \sin(q_n)-i/2}{u_j-2 g \sin(q_n)+i/2} &=&
\prod_{k\neq j}^{M}\frac{u_j-u_k-i}{u_j-u_k+i}\ \ \ \ \,\,\,\, \ \ \ \ \,\,\,\, , \,\,\,\, j=1,\dots,M \,, \la{HBAE}\,.
\eeqa
where  
\beq
g=-\frac{t}{U} \,. \nn
\eeq

In section \ref{generalwrap} we will review \cite{Hubbard} how, eliminating the electron momenta $q_n$ from these (twisted) Lieb-Wu equations, we obtain an effective set of (twisted) BAE for the spin degrees of freedom $u$ which are precisely the (twisted) BDS equations \cite{Beisert:2004hm}. In that section we will analyze wrapping corrections in full generality. We will see for example, that perturbatively in $g$ the effective (twisted) Bethe ansatz equations are valid up to order $g^{2L}$ for the choice of twists \cite{Hubbard} 
\beq
\phi_o-\phi_\updownarrow-\frac{\pi}{L}=0 \mod \frac{2\pi}{L}  \la{gal}
\eeq
and the $g^{2L}$ correction to any perturbative state will be given solely in terms of the position of the Bethe roots $u_j$ to leading ($g^0$) order only. This is probably a simplifying feature of the weak coupling finite size corrections which is likely to be present in the $\mathcal{N}=4$ supersymmetric spin chain. 

We will also study the generic case where (\ref{gal}) does not hold since it will provide us with a nice toy model to understand Luscher type corrections for many particle states.

In the following two sections we will consider a much simpler setup which however captures most of the relevant physical information. Namely we will consider  the very simple configuration with a single vacancy $o$ and a single double occupancy $\updownarrow$, that is $N=2$ and $M=1$.

\subsection{The magnon} \la{magnon}
A magnon in the Heisenberg XXX spin-$\frac{1}{2}$ chain, 
\beq
H_{xxx}= \frac{1}{4}\sum_n \(1- \vec{\sigma}_n \cdot \vec{\sigma}_{n+1} \) \la{Hxxx}
\eeq
is the lowest lying excitation above the ferromagnetic ground state.
It is a plane wave state 
$$
\sum_{n} e^{i pn} \sigma_{n}^- |\uparrow \dots \uparrow \rangle
$$
of one down spin in a chain of up spins, with excitation energy 
\beq
\epsilon(p)=1-\cos p \,.\la{xxx}
\eeq
In the Hubbard model such plane wave is not an eigenstate of the Hamiltonian but there is a close analogue of this state   when the empty site excitation $o$ and the double occupancy $\updownarrow$ form a bound state (note that a $o$ and a $\updownarrow$ in the same site is precisely the same as a spin down). More precisely, as reviewed in appendix A, we can consider the following half-filling state
\beqa
|\Psi\rangle=\sum_{n,n'}\psi(n,n') \(a^{\dagger}_{n,\updownarrow} a^\dagger_{n',o} \)  |\uparrow \dots\uparrow \rangle \,. \nn
\eeqa
with $\psi(n,n')$ being a superposition of plane waves with momenta $q$ and $q'$. By acting with the Hamiltonian on this state we can see that there exist bound state solutions with\footnote{Throughout this paper we always use this definition of $p$ which seems the most natural one from the Hubbard point of view. To make contact with the standard notations in the AdS/CFT literature we should use $p_{here}=p_{usual} +\pi$. } $q,q'=\frac{p}{2}\pm i \beta$ where $p$ is the bound state momentum while $\beta$ dictates the exponential damping of the wave function away from $n=n'$. The energy of such states, which we also call magnons, equals
\beq
\epsilon(p)=-4t\cos\frac{p}{2}\cosh\beta \la{Efin} \,.
\eeq
In infinite volume we find $\beta=\beta_\infty(p)$ where
\beq
4 g \cos\frac{p}{2}\sinh \beta_\infty(p)=-1 \la{betainf}
\eeq
so that $\epsilon(p)=\epsilon_\infty(p)$ with 
\beq
\epsilon_\infty(p)=-\sqrt{U^2+16t^2 \cos^2\frac{p}{2}} = -U\( 1+4g^2(\cos p +1)+\dots\)
\la{Eninf}
\eeq
which at weak coupling $g$ has the same $-\cos p$ dependence as (\ref{xxx}). 
This is  expected from the known result that perturbatively in small $g=-t/U$ the Hubbard model at half filling is a long-ranged Hamiltonian whose leading term is the Heisenberg spin chain. 

The magnon state can also be described by the triplet $(q,q^*,u)$ satisfying the Lieb-Wu equations 
\beqa
&& e^{i\(q-\phi_o \)L}=\frac{u-2 g \sin(q)-i/2}{u-2 g \sin(q)+i/2} \, , \ \ \ \ \ \ \  
e^{i\(q^*-\phi_o \)L}=\frac{u-2 g \sin(q^*)-i/2}{u-2 g \sin(q^*)+i/2} \,, \nn \\
&& e^{i (\phi_\updownarrow-  \phi_o)L }\frac{u-2 g \sin(q)-i/2}{u-2 g \sin(q)+i/2} 
\frac{u-2 g \sin(q^*)-i/2}{u-2 g \sin(q^*)+i/2} =1 \,, \nn
\eeqa
In infinite volume, the l.h.s of the first two equations is $0$ and $\infty$ for a bound state with complex momentum $q=\frac{p}{2}+i\beta$. This fixes 
\beq 
u= 2 g \sin q+\frac{i}{2} 
\,. \la{explode1}
\eeq
The reality of $u$ implies 
\beq
u=2 g \sin \frac{p}{2} \cosh \beta_\infty(p)\equiv u_\infty(p) \la{uuinf}
\eeq
and the condition (\ref{betainf}) which gives the dispersion relation (\ref{Eninf}).

When the state is put in finite volume, equation (\ref{Efin}) is still valid but the expression (\ref{betainf}) for $\beta(p)$ gets modified to 
\beq
4 g \cos\frac{p}{2}\sinh \beta=-\frac{\sinh\beta L}{\cosh\beta L -\cos\frac{L}{2}\(p-2\phi_o\)} \la{finitebeta}
\eeq
In appendix A we derive this equation from the direct study of the magnon wave function in finite volume.
Naturally, the same result can also be obtained from the Lieb-Wu equations (\ref{HBAE}).
Indeed these Bethe equations are exact for any $L$ since the interactions of the elementary particles are ultra local. 
The leading finite size correction to the magnon energy is then exponentially suppressed in the ratio of the system size $L$ by the the physical size $1/\beta_\infty$ of the bound state, 
\beq
\epsilon(p)-\epsilon_\infty(p)=U e^{-\beta_\infty(p)  L} \,2 \tanh\beta_\infty(p) \cos \frac{L}{2}\(p-2\phi_o \) 
\la{deltaElead}
\eeq
For the particular choice of twists (\ref{gal}), such that $e^{i(\phi_o-\phi_\updownarrow)L} =-1$, this leading correction vanishes, and instead we have
\beq
\epsilon(p)-\epsilon_\infty(p)=U e^{-2\beta_\infty(p)  L} \,2 \tanh\beta_\infty(p)
% \cos L\(p-2\phi_o \) 
\la{deltaEspecial}
\eeq

In the next section we will recover these known results from a diagrammatic point of view. This will turn out to be very useful to understand how to recover the finite size corrections from the effective theory point of view, that is, from the BDS language. In particular we will understand that if we were given solely the BDS Bethe equations we would not be able to recover the finite size corrections using a Luscher type approach \cite{Luscher:1985dn,Klassen:1990ub} except at strong coupling. Instead, the knowledge of the magnon constituents when computing the virtual processes wrapping the space-time cylinder will turn out to be essential. The diagrammatic language seems therefore promising to try to learn some lessons about what to expect for the $\mathcal{N}=4$ SYM chain if this chain most fundamental description comprises extra degrees of freedom \cite{Mann:2005ab,Hubbard,Gromov:2006dh,Gromov:2006cq,Gromov:2007cd,Rej:2007vm}.

\subsubsection{Diagrammatically} \la{diag}
\begin{figure} [t]   \centering {\includegraphics{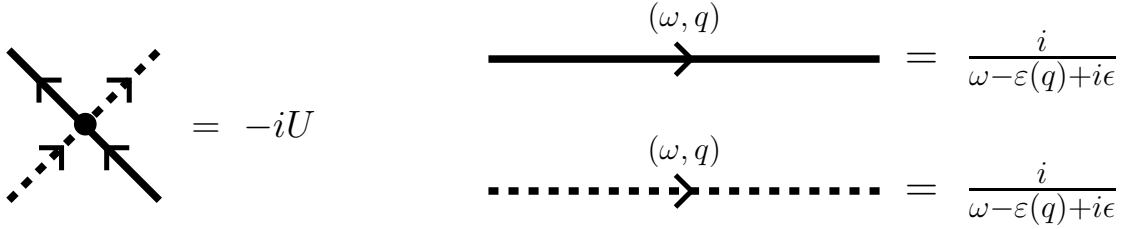}}
\caption{  Feynman rules for diagrammatic computations in the Hubbard model.
Each elementary particle ($o$ and $\updownarrow$) has a non-relativistic free propagator (solid and dashed lines) and there is only a quartic interaction vertex. Loops carry an extra minus sign as the elementary particles are fermions.
 }
\label{feynmanrules}
\end{figure}

In this section we shall explore the field theoretic description of the Hubbard model defined by the Hamiltonian (\ref{HHub}).
This will allow us to use the powerful diagrammatic techniques of field theory to obtain the finite size effects from loop diagrams with topological winding around the compact direction.
We start by writing the action of the theory 
$$
S=\int dt \sum_n \( \mathcal{L}_o +\mathcal{L}_{\updownarrow}+\mathcal{L}_{int} \)
$$
where the free part is given by
\beq
\mathcal{L}_{\sigma} = \frac{i}{2}\( a_{n,\sigma}^*\partial_t  a_{n,\sigma} -  a_{n,\sigma}\partial_ta_{n,\sigma}^*  \)
+t\(e^{i\phi_\sigma} a_{n,\sigma}^* a_{n+1,\sigma}  + e^{-i\phi_\sigma} a_{n,\sigma}^* a_{n-1,\sigma} \) \nn
\eeq
and the interaction by
\beq
\mathcal{L}_{int} =U  a^*_{n,o}a_{n,o}a^*_{n,\updownarrow}a_{n,\updownarrow}  \nn
\eeq
The elementary excitations $o$ and $\updownarrow$ have a non-relativistic propagator
\beq
\left\langle \,T\, a_{n,\sigma} (t) a^*_{n',\sigma}(t') \,\right\rangle=\int\frac{d\omega}{2\pi}
\int_{-\pi}^{\pi}\frac{dq}{2\pi} e^{-i\omega(t-t')+i(q-\phi_\sigma)(n-n')}
\frac{i}{\omega-\varepsilon(q)+i\epsilon} \nn
\eeq
where
\beq
\varepsilon(q)=-2t\cos(q)\,. \nn
\eeq 
The interaction term in the action gives rise to a quartic coupling with coupling constant $U$.
The Feynman rules are summarized in Figure \ref{feynmanrules}.
In a non-relativistic theory the number of particles is conserved and this greatly simplifies the field theoretic perturbative expansion of the theory \cite{Thacker:1974kv,Thacker:1976vp,Klose:2006dd}.
Diagrammatically, this fact stems from the retarded nature of the propagators which gives them an orientation.

In order to find the two particle spectrum we consider the two point function 
\beq
\left\langle \,T\, \chi_n(t) \chi_{n'}(t')\,\right\rangle=\int\frac{d\Omega}{2\pi}
\int_{-\pi}^{\pi}\frac{dp}{2\pi} e^{-i\Omega(t-t')+ip(n-n')}G\(\Omega,p\) \nn
\eeq
of the  composite operator
\beq
\chi_n(t)=a_{n,o} (t)a_{n,\updownarrow} (t)+a^*_{n,o} (t)a^*_{n,\updownarrow} (t) \nn
\eeq
In particular, to find out possible bound states (magnons) we should look for poles of $G(\Omega,p)$ thus obtaining the dispersion relation $\Omega(p)$.
Notice that there is a big arbitrariness in the choice of the composite operator $\chi$. The only requirement is that the state it creates has some overlap with the magnon wave function. 

\begin{figure} [t]   \centering  \resizebox{150mm}{!}{\includegraphics{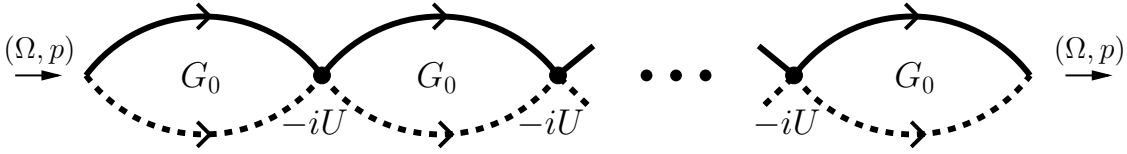}}
\caption{ 
Feynman diagrams for the two point function of the composite operator $\chi$.  }
\label{bubblediagrams}
\end{figure}
The propagator  $ G \( \Omega,p \) $ can be computed diagrammatically. It is given by the sum of all diagrams describing the two elementary particles moving freely and interacting $k$ times as shown in Figure \ref{bubblediagrams}.
More precisely, it is given by
\beq
G\(\Omega,p\)=\sum_{k=0}^{\infty} \left[ G_0\(\Omega,p\) \right]^{k+1} (-iU)^k
=\frac{i}{iG_0^{-1}\(\Omega,p\)-U} \la{GofG0}
\eeq
where the free propagator
\beqa
G_0\(\Omega,p\)=-\int\frac{d\omega}{2\pi}\int_{-\pi}^{\pi}\frac{dq}{2\pi} 
\frac{i}{\omega-\varepsilon(q)+i\epsilon}\frac{i}{\Omega-\omega-\varepsilon(p-q)+i\epsilon} \nn
\eeqa
can be computed by residues, 
\beqa
G_0\(\Omega,p\)
%&=&\int\frac{d\omega}{2\pi}\int_{-\pi}^{\pi}\frac{dq}{2\pi}
%\frac{i}{\omega-\varepsilon(q)+i\epsilon}\frac{i}{\Omega-\omega-\varepsilon(P-q)+i\epsilon}\\
%&=&\int_{-\pi}^{\pi}\frac{dq}{2\pi}
%\frac{i}{\Omega-\varepsilon(q)-\varepsilon(P-q)+i\epsilon}\\
%&=&\int_{-\pi}^{\pi}\frac{dq}{2\pi}
%\frac{i}{\Omega-4t+2t\( \cos(q) + \cos(q-P) \) +i\epsilon}\\
%&=&\int_{-\pi}^{\pi}\frac{dq}{2\pi}
%\frac{i}{\Omega-4t+4t \cos(q-\frac{P}{2})  
%\cos\frac{P}{2} +i\epsilon}\\
=\frac{i}{\sqrt{\Omega^2- 16t^2 \cos^2\frac{p}{2}    }}  \,. \nn
\eeqa
Thus, the full propagator reads
\beq
G\(\Omega,p\)=\frac{i}{\sqrt{\Omega^2 - 16t^2 \cos^2\frac{p}{2} } -U} \nn
\eeq
and the magnon bound state corresponds to the pole at
\beq
\Omega = -U\sqrt{1+16g^2 \cos^2\frac{p}{2}} \nn
\eeq
which is precisely the infinite volume result (\ref{Eninf}). 

With periodic boundary conditions of size $L$ the free propagator is changed.
The natural way to account for this effect is to sum over all possible windings of the loops in Figure \ref{bubblediagrams} around the compact circle. The winding numbers are a topological property of the Feynman graph that can be assigned to any of the propagators forming each loop. We shall compute the graph assigning the winding number $m$ always to the particle $o$. Furthermore, equation (\ref{GofG0}) remains valid provided 
\beqa
G_0\(\Omega,p\)&=&-\sum_{m=-\infty}^\infty\int\frac{d\omega}{2\pi}\int_{-\pi}^{\pi}\frac{dq}{2\pi}
\frac{i}{\omega-\varepsilon(q)+i\epsilon}
\frac{i}{\Omega-\omega-\varepsilon(p-q)+i\epsilon}e^{im(q-\phi_o)L} \nn \\
%&=&\sum_{m}\int_{-\pi}^{\pi}\frac{dq}{2\pi}
%\frac{ie^{im(q-\phi_o)L}}{\Omega-4t+4t \cos(q-\frac{P}{2})  
%\cos\frac{P}{2} +i\epsilon}\\
&=&\frac{i F_L\(\Omega,p\) }
{\sqrt{ \Omega^2-16t^2 \cos^2\frac{p}{2}     }} \nn
\eeqa
where,  if we parametrize $\Omega$ as 
\beq
\Omega=-4t \cos\frac{p}{2} \cosh \beta \la{Omegaofbeta}, 
\eeq
we get
\beq
F_L\(\Omega,p\)=1+ \sum_{m=1}^{\infty}e^{-mL\beta}\cos m L \(\frac{p}{2} -\phi_{o}\) 
=\frac{\sinh(\beta L)}{ \cosh(\beta L) - \cos L \(\frac{p}{2} -\phi_{o}\)} \nn
\eeq
The full propagator then becomes
\beq
G\(\Omega,p\)=\frac{iF_L\(\Omega,p\)}
{\sqrt{\Omega^2 - 16t^2 \cos^2\frac{p}{2} } -UF_L\(\Omega,p\)} \nn
\eeq
and the magnon pole sits at (\ref{Omegaofbeta}) with $\beta$ determined from the equation
\beq
4t \cos\frac{p}{2} \sinh \beta=U F_L\(\Omega,p\) \nn
\eeq
which is precisely (\ref{finitebeta}).

\subsection{Luscher in Hubbard} \la{Luscher}

Luscher developed a general formalism  \cite{Luscher:1985dn,Klassen:1990ub,Janik:2007wt}  to determine the leading finite size corrections in quantum field theory. In particular, he studied the change in the dispersion relation of one particle when one imposes periodic boundary conditions. Remarkably, he found that the leading correction to the energy $\epsilon(p)$ of a particle with momentum $p$ was completely fixed by the particles infinite volume dispersions relations and S-matrix. The idea is that the on-shell dispersion relation is defined by the pole of the propagator.
Following the standard notation of relativistic field theory we can write
\beq
iG^{-1}\(\Omega,p\)= -\Omega^2 + \epsilon^2_\infty(p)+\Sigma\(\Omega,p\)=0 \la{massshell}
\eeq
where 
$$
\Sigma\(\Omega,p\) =\Sigma_\infty\(\Omega,p\) +  \delta\Sigma\(\Omega,p\)
$$  
is the particle's self-energy, whose infinite volume part vanishes with zero derivative on the mass-shell $\Omega=\epsilon_\infty(p)$.
From (\ref{massshell}) with $\Omega=\epsilon_\infty(p)+\delta\epsilon(p)$ we obtain
\beq
\delta \epsilon(p)=\frac{\delta\Sigma }{2\epsilon_\infty}+\(\frac{\delta\Sigma}{2\epsilon_\infty}\)\( \frac{ \partial_0 \delta\Sigma}{2\epsilon_\infty}\)+\( \frac{1}{2} \partial^2_0 \Sigma -1\)
\frac{1}{2\epsilon_\infty} \(\frac{\delta\Sigma}{2\epsilon_\infty}\)^2  +\dots \la{gendeltaE}
\eeq
with all the quantities computed at the infinite volume mass-shell $\Omega=\epsilon_\infty(p)$.
The self-energy correction $\delta\Sigma$ can then be related to the S-matrix. 
This is achieved by evaluating the self-energy diagrams with winding around the compact direction
by deforming the integration over the loop momenta to pick the on-shell pole of the wound internal propagator 
and obtain an integral of the forward scattering S-matrix \cite{Luscher:1985dn,Klassen:1990ub,Janik:2007wt}.
More precisely, the leading finite size correction to $\epsilon^a(p)$ is given by\footnote{
In  \cite{Luscher:1985dn,Klassen:1990ub}  this formula was derived for relativistic theories and $p=0$. 
This amounts to computing the corrections to the particle's mass. 
In the process of derivation the propagator is wound around the spacial circle originating a factor of $\cos (q L)  $ multiplying the infinite volume propagator. Since the spacial momentum $p$ vanished the problem was isotropic with respect to the spacial directions and this factor could be simply replaced by $2 e^{i q L}$.  In the case $p\neq 0$ we  \textit{must} treat each exponential separately  which amounts to summing the contributions from the virtual particles going parallel and anti-parallel to the physical particle. In most cases, including those considered in this paper, these symmetrizations will simply lead to computing the real part of the result obtained keeping one exponential.
}
\beq
\frac{1}{2}\int_{\mathcal{C}} \frac{dq}{2\pi i} \,e^{i L q} \sum_b (-1)^{F_b}
 \(\epsilon_\infty^{b\,\prime} (q)-\epsilon_\infty^{a\, \prime}(p)\) \(1- S_{a\,b}^{a\,b}(p,q) \) +c.c.  \nn
\eeq
where $S_{a\,b}^{a\,b}(p,q)$ is the S-matrix for forward scattering of particle $a$ with particle $b$, $F_b=\pm 1$ encodes the bosonic/fermionic nature of particle $b$
and we consider only the contribution from diagrams with winding number $\pm 1$. For usual relativistic theories the contour $\mathcal{C}$ is given by an integral over the possible $S$-matrix poles plus an integral over the real axis. The latter describes a quantum loop and is absent in our case where the underlying theory is non-relativistic  \cite{Thacker:1974kv,Thacker:1976vp,Klose:2006dd}. The former can be simply computed by residues and is denoted by $\mu$-term.
\begin{figure} [t]   \centering 
 \resizebox{50mm}{!}
{\includegraphics{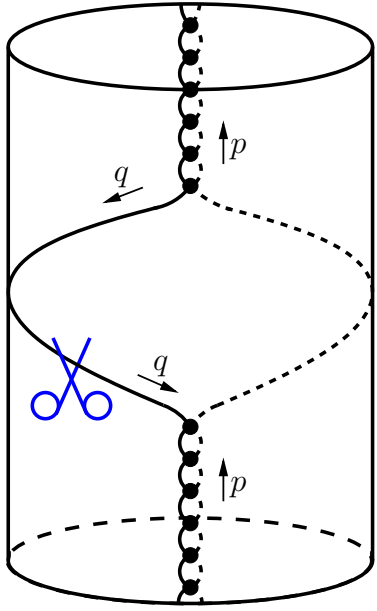}}
\caption{The leading finite size correction to the magnon energy is given by the Feynman diagram where its two elementary constituents split and merge after winding the spacetime cylinder once.
After cutting the wound loop, putting one elementary particle on shell, this diagram gives rise to the Luscher $\mu$-term.  }
\label{muterm}
\end{figure}
Let us now apply this general formalism to the Hubbard model and find the leading finite size correction to the magnon dispersion relation. As we saw in the previous sections the magnon is a bound state with a finite size $1/\beta_\infty$. Therefore, its energy in a finite system has a leading correction of order $e^{-\beta_\infty L}$, except for a particular choice of twists where this correction can be delayed to order $e^{-2\beta_\infty L}$. 
The diagram leading to the first 
wrapping correction corresponds to the splitting of the magnon into its fundamental constituents each one going around the space-time cylinder in opposite directions and meeting on the other side of the cylinder as depicted in figure \ref{muterm}. 
The Luscher $\mu$-term then reads 
\beq
\delta \epsilon(p)=\sum_{\sigma=o,\updownarrow} \( \frac{1}{2}\, {\rm Res}_{q} \,e^{i L(q-\phi_\sigma)}
 \(\varepsilon'(q)-\epsilon_\infty'(p)\) S_{\magnonexp,\sigma}^{\magnonexp,\sigma}(p,q) +c.c. \)\la{deltaepsilon}
\eeq
where $q$ and $ \varepsilon(q)$ are the momentum and energy of the elementary particle $o$ going around the loop in  figure \ref{muterm}, $ \epsilon_\infty(p)$ is the infinite volume dispersion relation of the magnon and the residue is taken at the pole of the $S$-matrix. 
The S-matrix between magnon and elementary particle ($o$ or $\updownarrow$) can be read of from the Bethe equations (\ref{HBAE})
\beq
S_{\magnonexp,o}^{\magnonexp,o}(p,q)=S_{\magnonexp,\updownarrow}^{\magnonexp,\updownarrow}(p,q)=\frac{u_\infty(p)-2 g \sin(q)-i/2}{u_\infty(p)-2 g \sin(q)+i/2} \nn
\eeq
with
\beq
u_\infty(p)= 2g \sin \frac{p}{2} \cosh \beta_\infty(p)  \nn \,.
\eeq
The pole condition
\beq
 2 g \sin(q)=u_\infty(p)+i/2 \nn
\eeq
has the simple solution 
\beq
q=\frac{p}{2}+i\beta_\infty(p) \nn\,.
\eeq
Putting everything together we recover the result (\ref{deltaElead}).
For the choice of twists (\ref{gal}) this leading term vanishes and one needs to consider the contribution coming from diagrams with winding number $\pm 2$ \footnote{For generic twists, the subleading correction (of order $e^{-2\beta_\infty L}$) to the magnon dispersion relation depends on all terms in (\ref{gendeltaE}), including  the second derivative of the infinite volume self-energy which is not a physical on-shell observable. 
However if $\delta \Sigma$ vanishes at order $e^{-\beta_\infty L}$ then we only need to compute the first term, that is, the correction to the self energy due to winding 2.},
\beq
\delta \epsilon(p)= {\rm Res}_{q} \,e^{i 2 L(q-\phi_o)}
 \(\varepsilon'(q)-\epsilon_\infty'(p)\) S_{\magnonexp,o}^{\magnonexp,o}(p,q) +c.c.  \nn
\eeq
which gives (\ref{deltaEspecial}).

\subsubsection{Luscher with magnon-magnon S-matrix}
In the previous sections we saw, by many means including a Luscher type computation, that the leading finite size correction to the magnon dispersion relation is given by \cite{Hubbard}
\beq
\delta \epsilon(p)=U e^{-2\beta_\infty  L} \,2 \tanh\beta_\infty  \nn
\eeq
when the Hubbard twists are chosen as in (\ref{gal}).
%We also used the fact that the momenta is quantized in such a way that the cosine in (\ref{}) can be replaced by $1$.
This expression is valid for all values of the coupling $g$. In particular, at strong coupling we find \cite{Astolfi:2007uz}
\beq
\delta \epsilon(p)= 
\(-\frac{U}{2g}{\sec\frac{p}{2}} +\frac{U}{64g^3}\sec^3\frac{p}{2} +\dots \)
\exp{- \frac{L}{2g} \sec\frac{p}{2} \(1-\frac{1}{96g} \sec^2\frac{p}{2} +\dots \)}  \,.  \la{exactstrong}
%\delta \epsilon(p)= U e^{-\frac{L}{2g \cos\frac{p}{2}}} \(\frac{1}{2g}{\sec\frac{p}{2}} +\frac{1}{384g^3}\sec^3\frac{p}{2}\(\frac{L}{g}\sec\frac{p}{2}-6\)+\mathcal{O}(1/g^5)\)
\eeq
In \cite{Janik:2007wt} -- in the process of computing the leading $\mu$-term prefactor for the $AdS_5\times S^5$ giant magnon -- Janik and Lukowski also determined what the contribution would be for the BDS scenario. They found 
\beq
\delta \epsilon(p)= -\frac{U}{g}{\sec\frac{p}{2}} e^{-\frac{L}{2g} \sec\frac{p}{2}}  +\dots\nn
\eeq
where we adapted their results to our normalizations. 
This result was obtained applying Luscher formula with the magnon-magnon S-matrix and thus describes a different physical process from the point of view of the elementary particles. Surprisingly, the two results agree up to an overall factor of 2.
Physically one can try to justify this result using the fact that for large $g$ the magnon is a weakly bound state of the elementary particles $o$ and $\updownarrow$. Thus, it is not unreasonable that the effect of a magnon loop around the spacetime cylinder is almost the same as a loop of the elementary particles with winding number 2.
In any case, the accidental nature of this agreement is confirmed by its limitation to the strong coupling regime. 
In appendix C we repeat the computation of \cite{Janik:2007wt}, expanding the result further in $1/g$,

\beq
\delta\epsilon(p)=\Re \(-\frac{U}{g} \sec \frac{p}{2} +  \frac{iU}{g^2}\tan\frac{p}{2} \sec^2\frac{p}{2} 
%-\frac{1}{128g^ 3} \( 23 \cos p -49\) \sec^5\frac{p}{2}
+\dots \) 
\exp -\frac{L}{2g}\sec\frac{p}{2}\(1 -\frac{i}{2g}\tan\frac{p}{2} \sec\frac{p}{2} +\dots \)\,. \la{lumagmag}
\eeq
So that the result starts differing (apart from the factor of $2$) at the next order in $1/g$. If we fix $L/g$ and expand both the prefactor and exponent then we can easily see that the $1/g^2$ term drops out and the mismatch is delayed to the next to leading  order in $1/g$, as mentioned in the introduction. This limit is the one usually considered in the AdS/CFT context.

The BDS magnon effective theory, reviewed in the next section, contains bound states of magnons \cite{Dorey:2006dq} which are simple generalizations of the Bethe strings in the Heisenberg model. In \cite{Hatsuda:2008gd} the Luscher term accounting for the finite size correction to the (dyonic) Giant magnon \cite{Dorey:2006dq} dispersion relation was written including the virtual exchange of all these bound states of magnons. Thus, one might question if the sum of the contributions of all these bound states will correct the result (\ref{lumagmag}) to give the exact result (\ref{exactstrong}). This is \textit{not} the case. 
In fact all diagrams present in the theory are those in figure \ref{bubblediagrams} and they do {\it not}, in any sense, describe virtual exchanges of magnons and magnon bound states. At most, at very large $g$ the loops of $b$ magnons could be mimicking $2b$ loops of fundamental particles and thus, the best we can expect from summing all the possible magnons is to recover the \textit{full finite volume result} but always for $g \to \infty$.

Furthermore, in the weak coupling regime the magnon-magnon computation gives the right $g^{2L}$ coupling dependence but misses completely the momentum dependence of $ \delta \epsilon$. 

This result puts in question the validity of direct application of Luscher formulas to the giant magnon of $\mathcal{N}=4$ SYM. In particular, if the giant magnon is not an elementary particle (like the 2 electrons bound state in the Hubbard model) of the worldsheet theory then its finite volume energy will be sensitive to the elementary excitations of the theory and can not be recovered just from the magnon-magnon S-matrix in infinite volume.
On the positive side, this makes the wrapping effects a window into the most elementary level of the theory.

\subsection{Magnon effective Bethe Ansatz equations and generic wrappings} \la{generalwrap}
In this section we will review the results of \cite{Hubbard} and see how the Bethe equations for the effective system of spins at half-filling coincides with the (twisted) BDS equations. We will then study the wrapping interactions for a general state. In particular we will understand how to \textit{dress} the BDS equations in such a way that they include the wrapping interactions. We will also see that  the leading order correction to the energy of a generic $M$ magnon state can be given a Luscher type diagrammatic interpretation. 

At half filling the Bethe ansatz equations (\ref{HBAE}) are given by
\beqa
&&e^{i\(q_n-\phi_o \)L}=
\prod_{j=1}^{M}\frac{u_j-2 g \sin(q_n)-i/2}{u_j-2 g \sin(q_n)+i/2} \, ,  \la{bae1}\\
&&e^{i\(q_{n+M}-\phi_o \)L}=
\prod_{j=1}^{M}\frac{u_j-2 g \sin(q_{n+M})-i/2}{u_j-2 g \sin(q_{n+M})+i/2} \,, \la{bae2}\\
&&e^{i (\phi_\updownarrow-  \phi_o)L }\prod_{n=1}^{M} \frac{u_j-2 g \sin(q_n)-i/2}{u_j-2 g \sin(q_n)+i/2} \frac{u_j-2 g \sin(q_{M+n})-i/2}{u_j-2 g \sin(q_{M+n})+i/2} =
\prod_{k\neq j}^{M}\frac{u_j-u_k-i}{u_j-u_k+i} \la{bae3}\,,
\eeqa
where we explicitly split the $2M$ momenta $q_n$ into two equal groups so that all indices range from $1$ to $M$.  A state with $M$ magnons is a state where the $3M$ Bethe roots organize in triplets of Takahashi states
\beq
(q_n,q_{n+M},u_n) \,, \la{Takahashi}
\eeq
with
\beq
q_{n+M}=q_n^* \,, \nn
\eeq
while the real $u$ root is given by
\beq 
u_n\simeq 2 g \sin q_n+\frac{i}{2} = 2 g \sin q_n^*-\frac{i}{2}  \,. \la{explode}
\eeq
The last condition is necessary once we allow $q_n$ to have a positive imaginary part (and thus $q_{n+M}=q_n^*$ to have a negative imaginary part). In this case, the l.h.s of equations (\ref{bae1}) and (\ref{bae2}), respectively,  vanishes and diverges exponentially with the system size $L$ and condition (\ref{explode}) is required to obtain the same behavior for the  r.h.s.  It is also clear - and shown in \cite{Hubbard} - that if we multiply the equations for each element of the Takahashi triplet we obtain the twisted BDS equations  
\beqa
e^{i (p_n-\phi_o-\phi_\updownarrow  )L}  \simeq  \prod_{m\neq n}^{M}  \frac{u_n-u_m+i}{u_n-u_m-i}\,, \la{BDS}
\eeqa
parametrizing the momenta as 
\beq
q_n=\frac{p_n}{2} + i \beta_n \,. \nn
\eeq
The dispersion relation $\epsilon(q_n)+\epsilon(q_{n+M})$ as function of $p$ and the relation between $p$ and the Bethe roots $u$ is to leading order exactly as before, as it follows simply from computing the real and imaginary part of (\ref{explode}), namely $\beta_n  \simeq \beta_\infty(p_n)$ given in (\ref{betainf}), 
\beq
u_n\simeq u_\infty(p_n)\equiv  \frac{1}{2} \tan\frac{p_n}{2} \sqrt{1+16g^2 \cos^2\frac{p_n}{2}} \la{uinf} \,,
\eeq
and $E\simeq \sum \epsilon_\infty(p_n)$ with 
\beq
\epsilon_\infty(p)\equiv -\sqrt{U^2+16t^2 \cos^2\frac{p}{2}} \la{einf} \,.
\eeq
\subsubsection{Corrected BAE} \la{corrected}

In this section we will study expressions (\ref{BDS},\ref{uinf},\ref{einf}) in greater detail. That is we will understand how these relations get modified once the leading finite size effects are taken into account. To do so we will study the leading wrapping effects for a generic many magnon state. 

Physically there are two sources of corrections to the energy of a many magnon state. On the one hand, the energy of the state as a function of the magnon momenta  changes when the state is put in finite volume. This will lead to a Luscher type correction. On the other hand, the periodicity condition, that is the BAE, are corrected and thus the quantized momenta are shifted. Diagrammatically this last effect would be due to new wrapping virtual processes correcting the magnon S-matrix rather than to the usual self energy virtual graphs present for the Luscher type contribution.

As for the single magnon case we will see that the bound state structure of the magnon must  be taken into account to reproduce the proper finite volume results.

Needless to say, instead of correcting the effective BDS equations we could simply use the exact Lieb-Wu nested Bethe equations! The point is that we want to understand how the corrections to the Bethe equations of effective spin theories look like. This might be useful if, as discussed in the introduction, the $\mathcal{N}=4$ SYM long-ranged Hamiltonian stems  from an underlying Hubbard like description.

%In small $g$ perturbation theory, we will see that the leading corrections appear at order $g^{L}$ for generic twists and at order $g^{2L}$ if the twists are carefully chosen as in (\ref{gal}). We will also see that, in order to compute the leading corrections for a generic state whose Bethe roots were computed from the uncorrected (twisted) BDS equations up to wrapping order, we need only know the $g^0$ order position of these Bethe roots! 

We want to eliminate the magnon momenta $q_n$ from the Lieb-Wu equations thus obtaining an effective equation for the magnon rapidities $u_n$. As we saw, to leading order we simply have (\ref{explode}) but since we want also to keep track of the leading wrapping corrections to the effective equations we should instead write $q_n=\frac{p_n}{2}+i\beta_n$ and
\beq
u_n=2 g \sin q_n+\frac{i}{2}+\Delta_n=2 g \sin q_n^*-\frac{i}{2}+\Delta_n^* \la{uD}
\eeq
where $\Delta_n$ is a small quantity which can be computed from the equations (\ref{bae1}) and (\ref{bae2}) for the magnon momenta. This is done in appendix B.
Since $u_n$ is real we can compute both $u_n$ and $\beta_n$  from the knowledge of the real and imaginary of the small quantity  $\Delta_n$. 
In particular, taking the imaginary part of (\ref{uD}), we obtain that $\delta \beta_n \equiv \beta_n-\beta_\infty(p_n)$ is given by
\beq
\la{deltabeta} \delta\beta_n \simeq   2 \tanh\beta_\infty(p_n)\, \Im\(\Delta_n\)  
% =-\frac{i\(\Delta_n-\Delta_n^*\) }{4g \sin \frac{p_n}{2} \cosh\beta_{\infty(p_n)}}
\eeq
where as before $\beta_\infty(p)$ is defined through (\ref{betainf}).
To proceed we introduce the notation $q_n^\infty \equiv \frac{p_n}{2}+i\beta_\infty(p_n)$ and $u^\infty_n \equiv2 g \sin q^\infty_n-\frac{i}{2}$ so that
\beq
u_n = u_n^\infty+2 g\, i \cos q_n^\infty \,\delta \beta_n+ \Delta_n + \mathcal{O}(e^{-2\beta L}) \,, \la{uper}
\eeq

Next, as explained in greater detail in appendix B, we multiply the three Lieb-Wu Bethe equations for the Takahashi triplet (\ref{Takahashi}) and expand using (\ref{uD}) and (\ref{uper}) to find
\beq
e^{i (p_n-\phi_\updownarrow-\phi_o)L} = \prod_{m\neq n} \frac{u^\infty_n-u^\infty_m+i}{u^\infty_n-u^\infty_m-i} \,e^{i \phi_{nm}} +\mathcal{O}(e^{-2\beta L})  \,, \la{BDSupgrade}
\eeq
where the \textit{wrapping dressing kernel} $\phi_{nm}$ is given by\footnote{If the twists are chosen as in (\ref{gal}) the imaginary part of $\Delta_n$ becomes of order $e^{-2\beta L}$ and we need to expand further. In this case we obtain (\ref{phispecial}) in Appendix B
and (\ref{uper},\ref{BDSupgrade}) hold to order $e^{-3 \beta L}$. 
}
\beq
\phi_{nm}=-\frac{\Im(\Delta_n)}{(u_n^\infty-u_m^\infty)^2+1}\(\frac{1}{u_n^\infty} \tan^2\frac{p_n}{2}+\frac{2}{u_n^\infty-u_m^\infty}\) - (n\leftrightarrow m)  \la{dressing}
\eeq

The dressed Bethe equations (\ref{BDSupgrade}) can trivially be solved perturbatively provided a solution $p_n^{\infty}$ to the original BDS equations is given. In this case $u^\infty_n$, which was a \textit{function} of $p_n$, can be expanded around the value $p_n=p_n^\infty$. The value of the Bethe roots for these values of momenta are denoted by ${\bf  u^{\infty}_n}$. We stress again, the ${\bf u_n^\infty}$ are the values of the Bethe roots obtained via the BDS equations whereas $u_n^\infty$ are \textit{functions} of a free variable $p_n$. Writing $u^{\infty}_n={\bf u^\infty_n}+\delta u_n$ we easily see that the leading order shifts to the Bethe roots due to the inclusion of the dressing Kernel reduces to the simple linear problem
\beq
 L \delta p_n
 -\sum_m\frac{2 \(\delta u_m-\delta u_n\) }{({\bf u_n^{\infty}}-{\bf u_m^{\infty}})^2+1}   = \sum_{m\neq n}\phi_{n,m}  \nn
 \eeq
with $ \delta u_n= \(\frac{d\,u^\infty_n }{d\,p_n}\)_{p_n=p_n^\infty} \delta p_n $.  Having found $\delta p_n$ and $\delta \beta_n$ we want to compute the shift to the many particle state energy 
\beqa
\delta E=\sum_n -4t \cos\frac{p_n}{2}\cosh \beta_n+4t\sin\frac{p_n^\infty}{2}\cosh \beta_\infty(p_n^\infty) \nn 
\eeqa
due to the wrapping effects. As mentioned above, this expression is non-zero due to two completely distinct type of corrections. 

On the one hand, as we just saw, the momenta are quantized differently and thus we will have a contribution due to the displacement of the Bethe roots when wrapping interactions are taken into account. For many particles states like the ones we are now considering these corrections must be taken into account. 

On the other hand the functional dependence of the energy on the momenta $\{p_n\}$ changes when we put the system in finite volume. More precisely $\beta_n$ will be given by the infinite value expression $\beta_\infty(p_n)$ plus the correction $\delta \beta_n$  which will in general depend on all the magnon momenta in an entangled way. This contribution is precisely the analogue of the Luscher corrections described in the previous sections for the single magnon case. We will discuss these corrections in greater detail in the next subsection. 

Putting these two corrections together we immediately get 
\beq
\delta E=\sum_n \(-U \delta \beta_n + \frac{d \epsilon_\infty(p_n)}{dp_n} \delta p_n\)  \la{deltaE} \,.
\eeq
Roughly speaking we could say that the first term is of Luscher type and accounts for virtual processes correcting the magnon dispersion relations while the second term stems from correcting the magnon $S$-matrix and therefore the BAE.  It would be interesting to provide $\phi_{nm}$ with a diagrammatic interpretation. Moreover, in the thermodynamical Bethe ansatz approach to the computation of finite size effects in relativistic integrable models, \textit{renormalized} Bethe equations for the positions of extra zeros in the TBA $Y$-system appear \cite{Bazhanov:1996aq}. If the same structure emerges for the AdS/CFT TBA equations then the form of the above dressed equations (\ref{BDSupgrade}) might provide some hints about the possible aspect of such dressed equations.

\subsubsection{Meeting (and generalizing) Luscher} 
\begin{figure} [t]   \centering 
 \resizebox{50mm}{!}
{\includegraphics{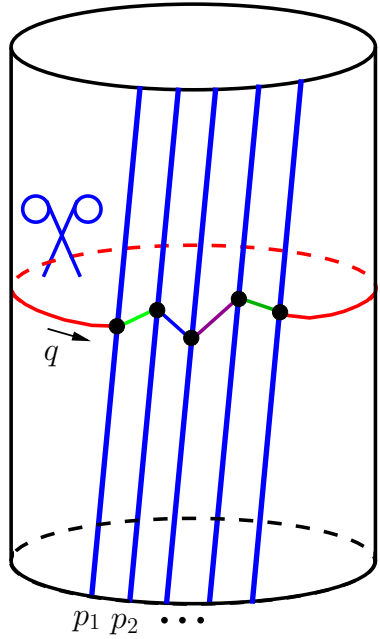}}
\caption{ 
The many particle state is corrected due to interactions with a virtual particle going around the spacetime cylinder. For integrable theories the correction to the energy of the state is expressed in terms of a product of factorized scattering matrices between the virtual particle and the various physical particles. For non-diagonal scattering this product defines a transfer matrix, a central object in quantum integrability.}
\label{fig:manyLuscher}
\end{figure}

\la{genluscher}
In this section we want to analyze the first correction $\delta E_{Luscher}=-U\sum_{n=1}^M \delta \beta_n$ to the energy of a M magnon state when put in finite volume and provide it with a simple physical diagrammatic meaning\footnote{This section, together with appendix D benefited largely from discussions with K. Zarembo.}. Using (\ref{deltabeta}) and the expression for $\Delta_n$ in appendix B we find  
\beqa
\delta E_{Luscher} = \frac{U}{2}\sum_{n=1}^{M} 2 \tanh \beta_\infty(p_n)\,e^{i\(\frac{p_n}{2}+i\beta_\infty(p_n) -\phi_o \)L}\prod_{m\neq n}^{M} 
\frac{{ \bf u^\infty_m}-{\bf u^\infty_n} +i}{{\bf u^\infty_m}-{\bf u^\infty_n}}  +c.c. \nn
\eeqa 
which can be written as 
\beq
\delta E_{Luscher} =\frac{1}{2}\sum_{n=1}^M \sum_{\sigma=o,\updownarrow} \int_{\mathcal{C}_n} \frac{dq}{2\pi i} \(\varepsilon'(q)-\epsilon_\infty'(p_n)\)\,e^{i \(q -\phi_o \)L}\prod_{m=1}^{M} 
S_{\magnonexp,\sigma}^{\magnonexp,\sigma}(p_m,q) + c.c. \la{LuscherM}
\eeq 
where $\mathcal{C}_n$ encircles the pole of $S_{\magnonexp,\sigma}^{\magnonexp,\sigma}(p_n,q)$ at $q=\frac{p_n}{2}+i\beta_\infty(p_n) $ in the counter-clockwise direction. Obviously this expression resembles the single magnon Luscher formula (\ref{deltaepsilon}) used before  and can be thought of as its many particle generalization. Physically it represents the correction to the state self-energy due to the process where a virtual particle with momenta $q$ goes around the cylinder scattering with all other physical excitations as depicted in figure \ref{fig:manyLuscher}.

Due to its strikingly simple form one might try to guess what the many particle Luscher formula for generic quantum integrable two dimensional field theories with factorized scattering could be. A likely candidate for such expression 
for a state with $M$ particles with polarizations $a_1,\dots, a_M$ and momenta $p_1,\dots, p_M$
would be 
\beqa
\delta E_{Luscher} &=&\Re\Big\{ \sum_{n=1}^M \sum_{\{b_1,\dots,b_M\}} \int  \frac{dq}{2\pi i} \(\varepsilon_{b_n}^{\prime}(q)-\varepsilon_{a_n}^{\prime}(p_n)\)\,e^{i \(q -\phi_{b_1} \)L} (-1)^{F_{b_1}} \la{conjecture}\\
&&\left. \,\,\, \,\,\, \,\,\, \[S_{a_1, b_1}^{a_1,b_2}(p_1,q)S_{a_2, b_2}^{a_1,b_3}(p_2,q)\dots S_{a_M, b_M}^{a_M,b_1}(p_2,q) - \delta_{b_1}^{b_2} \dots \delta_{b_M}^{b_1} \] \right\} \nn
\eeqa
where we sum over the fundamental polarization $b_j$ but in principle allow the physical particle to be bound states in which case some of the indices $a_n$ will be bound states indices as in (\ref{LuscherM}). In this case, the corresponding S-matrices in this expression should be the usual fused $S$-matrices. For fermionic virtual particle we include the standard $-1$ factor from the loop and in case magnetic fields are coupled to any of the particles a corresponding twist is included. Notice that we allow the polarization of the virtual particle to change as it scatters with each of the physical particles. This is not in contradiction with (\ref{LuscherM}) because due to charge conservation $S_{\magnonexp,\sigma}^{\magnonexp,\sigma'}$=0 if $\sigma\neq \sigma'$. We also notice that, not surprisingly, the first term in the second line can be written in a very compact form in terms of the transfer matrix $\hat T(q)={\rm str}_0 \(\hat S_{1,0}(p_1,q)\dots \hat S_{M,0}(p_M,q)\)$, a central object in integrable theories. Finally the second term in the last line is irrelevant if we only integrate over the S-matrices poles but for relativistic theories we expect the forward scattering amplitude to appear when the momenta is integrated over the real axis. 

In Appendix $D$ we consider a direct application of this formula for the $SO(\infty)$ sigma model. An independent computation of the results therein presented would be useful  to put the conjecture (\ref{conjecture}) in firmer ground.

In the AdS/CFT setup it would be interesting to consider this expression applied to the computation of the exponential corrections to spinning strings \cite{SchaferNameki:2006gk,SchaferNameki:2006ey}. In the scaling limit the transfer matrix becomes the exponential of the algebraic curve quasimomenta \cite{Kazakov:2004qf,Beisert:2005bm,Beisert:2004ag,SchaferNameki:2004ik,Beisert:2005di}� and using the techniques in \cite{Gromov:2007aq,Gromov:2007cd,Gromov:2007ky,Vicedo:2008jy,Gromov:2008ie,GSVfuture} one might try to study semi-classical wrapping corrections around fairly general classical solutions. 

Finally, we should stress that expression (\ref{conjecture}) is a conjecture for which we have no prove but only empirical evidence. It would be interesting to try to directly generalize Luscher arguments for many particles states in integrable theories in which factorizability should provide dramatic simplifications.

\subsubsection{Perturbative treatment} \la{BDSper}
In this section we will consider the perturbative small $g$ regime. Since 
\beq
e^{-\beta_\infty}=2g \cos\frac{p}{2} +\mathcal{O}(g^3) \nn
\eeq
the leading finite size corrections for generic twists will appear at order $g^L$ and at order $g^{2L}$ for twists given by (\ref{gal}). For example the single magnon results of section \ref{magnon} are easily seen to give
\beqa
\epsilon(p)-\epsilon_\infty(p)&\simeq &\left\{ 
\begin{array}{ll}
\displaystyle -2 U\(2g\cos \frac{p}{2}\)^{L}  \cos\frac{L}{2}(p-\phi_o) &,  \text{for generic twists}  \\ \\
\displaystyle -2 U \(2g\cos\frac{p}{2}\)^{2L} &,  \text{for twists as in (\ref{gal}) }
\end{array} \la{prediction}
\right.
\eeqa
As explained in section \ref{corrected}, when we want to consider states with more than one magnon we expect two types of contributions. On the one hand, we have a Luscher type contribution of the form
\beq
\delta E_{Luscher}\simeq - U \sum_{n=1}^{M} \delta \beta_n \la{LuscherE}
\eeq
due to the fact that the energy of the state, as a function of the magnon momenta, changes when the state is put in finite volume. On the other hand we obtain extra corrections due to the fact that the effective Bethe equation for the magnons are corrected when wrapping effects are taken into account, 
\beq
\delta E_{BAE}\simeq  \sum_{n=1}^{M}  \frac{d \epsilon_\infty(p_n)}{dp_n} \delta p_n\, . \nn
\eeq
However, since perturbatively the derivative of the dispersion relation with respect to $p$ carries an extra  power of $g^2$, (\ref{LuscherE}) is sufficient to compute the leading wrapping correction. Notice that since the exponential factors of $e^{-\beta_\infty L}$ start at $g^L$ order the prefactors can be computed using the $g^0$ order positions of the Bethe roots $u_n^\infty$ (which we simply denote by $u_n$ in what follows).  This is a huge simplification which is probably also present in $\mathcal{N}=4$ SYM if the most fundamental description of the supersymmetric chain bears a resemblance with the Hubbard model.

Generic states can be studied by expanding at small $g$ the expression in Appendix B. For concreteness let us focus on $2$ magnon states with twists given by (\ref{gal}). Moreover we chose the twists as given in \cite{Hubbard} which correspond not only to (\ref{gal}) but also to $e^{i\phi_\updownarrow L}=(-1)^{L+1}$. In this case not only the wrapping order is delayed to $g^{2L}$ but also the BDS effective equations become untwisted. We obtain, to leading $g^{2L}$ order,
\beqa
\frac{\delta E}{U} =  2 \( \Delta_1^{(0)}\)^2+2 \( \Delta_2^{(0)}\)^2 + \frac{4 \,\Delta_1^{(0)} \Delta_2^{(0)}}{(u_1-u_2)^2+1} +o(g^{2L}) \la{finalper}
\eeqa
with
\beqa
\Delta_{1,2}^{(0)}= i g^L \(\frac{i}{u_{1,2}-i/2}\)^L \frac{u_{1,2}-u_{2,1}-i}{u_{1,2}-u_{2,1}}  \,. \nn
\eeqa
For Konishi like states with opposite momenta $p$ and $-p$ (and thus $u_1=-u_2$) we get the remarkably simple expression  
\beqa
\frac{\delta E}{U}=2^{2L+1}g^{2L} (\cos(p)+3)\csc^2\frac{p}{2}\cos^{2L}\frac{p}{2}  \,. \la{finalper2}
\eeqa
In particular, we might consider some "\textit{high energy magnons}" with $p \simeq -\pi $ in a long spin chain to find 
\beqa
\delta E^{\text{large momentum}}  \sim (p+\pi)^{2L} \ll 1 \nn
\eeqa
which is a tiny quantity while for low momentum states with $p\simeq 0 $ 
\beqa
\delta E^{\text{low momentum}}   \sim \frac{1}{p^2}  \gg 1\,, \nn
\eeqa
and wrapping corrections are very large. This is physically intuitive as low momentum states probe larger portions of the space cylinder and thus are more sensitive to wrapping interactions. 

It would be very interesting if the computations of \cite{w1,w2} could be generalized to generic two magnon states with arbitrary $L$. If this turns out to be feasible then the  study of the dependence of the anomalous dimensions on the magnon momenta could be an important window 
into the structure of wrapping effects in $\mathcal{N}=4$ SYM.

As mentioned above, to compute (\ref{finalper}) or (\ref{finalper2}) we only need to know the leading $g^0$ position of the Bethe roots which are simply given by  the Heisenberg chain Bethe equations. If $u_1=-u_2=u$ they become simply
\beq
\(\frac{u+i/2}{u-i/2}\)^L=\frac{u-(-u)+i}{u-(-u)-i} \nn
\eeq
so that the only effect of the second magnon is to effectively renormalize $L$ to $L-1$. Therefore the momenta $p=2 \arctan 2u $ will be simply given by  $p=-\pi+\frac{2\pi n}{L-1}$. E.g. for the Konishi operator $p=-\frac{\pi}{3}$ and $\delta E= -2268 g^8$ which is represented by the mismatch between curves $c_4$ and $c_2$ in figure \ref{fig1}. 

Another example which illustrates the huge variance of the prefactor as function of the magnon momenta is the $2$ magnon state for some large chain with, e.g.,  $L=100$. For these states the wrapping corrections range from the smallest values for the largest momentum states with $n=49$ for which $\delta E\simeq -7.97 \times 10^{-300} g^{200}$ to the highest value for the lowest momentum states with $n=1$ and corresponding $\delta E=-1.15 \times 10^{64} g^{200}$.

\section{Families of long-ranged integrable Hamiltonians} \la{Toy2}

In the previous sections, we explored wrapping effects in the Hubbard model, as a controlled toy model closely related to
$\mathcal{N}=4$, see discussion in the introduction.
In that model the long range Bethe equations for the magnons are effective equations and the fundamental degrees of freedom are electrons whose interactions are ultra local. 
In this section, we will explore a completely different toy model which also has an analytic solution. In this model, the fundamental description is given by a long ranged Hamiltonian where,  by construction, wrapping interactions are under control. Contrary to the previous model,  it does not seem to share many features with the known $\mathcal{N}=4$ spin chain however it is a nice simple model worth looking at. 

The algebraic Bethe ansatz construction is the formalism that diagonalizes transfer matrix operators like (\ref{Tintro}) mentioned in the introduction\footnote{A transfer matrix also appeared in section \ref{genluscher} although there, the operators being multiplied inside the trace were $S$-matrices rather than $R$-matrices.}. These transfer matrices commute with one another for different values of the spectral parameter\footnote{The reason for this is that the operators being multiplied in (\ref{Tintro}) or in (\ref{Tsu2}) below obey the Yang-Baxter triangle relation.}.
Thus, if we construct a spin chain Hamiltonian $H$ from the transfer matrix (usually, by taking derivatives of its logarithm  at a particular point $u^*$) then, by construction $\[H,T(u)\]=0$ and we immediately obtain a huge number of conserved charges and hence quantum integrability.

For example, let us consider the standard  $SU(2)$ spin chain transfer matrix 
%\beq
%\hat T(\lambda)\equiv {\rm tr}_{0} \(\prod_{j=1}^{L} \(\lambda-\frac{i}{2}\)+i\mathcal{P}_{0j} \) \,, \la{Tsu2}
%\eeq
\beq
\hat T(\lambda)\equiv {\rm tr}_{0} \(\prod_{j=1}^{L} \, \frac{\lambda+i\, \mathcal{P}_{0j}}{\lambda+i} \) \,, \la{Tsu2}
\eeq
where $\mathcal{P}_{0j}$ is the permutation operator between a physical vector space $h_j$ and an auxiliary space $h_0$, both isomorphic to $\mathbb{C}^2$. The transfer matrix is then an operator acting in the full Hilbert space given by the tensor product of $L$ copies $h_1,\dots, h_L$ associated with the $L$ spin chain sites.
The eigenvalues of this $SU(2)$ transfer matrix are given by
\beq
T(\lambda)=  \prod_{j=1}^{M}\frac{\lambda-u_j-i/2}{\lambda-u_j+i/2}+\(\frac{\lambda}{\lambda+i}\)^L \prod_{j=1}^{M}\frac{\lambda-u_j+3i/2}{\lambda-u_j+i/2} \la{Teigen} \,
\eeq
where $u_j$ are denoted by Bethe roots. It is clear from the definition (\ref{Tsu2}) that the eigenvalues can not have poles as $\lambda$ approaches $u_j-i/2$. The cancellation of the corresponding residues in (\ref{Teigen}) is indeed guaranteed by the Bethe equations
\beq
\(\frac{u_j+i/2}{u_j-i/2}\)^L=\prod_{k\neq j}^{M}\frac{u_j-u_k+i}{u_j-u_k-i} \, \la{BAEsu2}
\eeq
which quantize the Bethe roots $u_j$. Physically, the left hand side of this equation represents the free propagation $e^{ip_j L}$ of the magnon $j$ as it goes around the spin chain while the r.h.s. $\prod_{k\neq j} S(p_j,p_k)$ describes the scattering of this magnon with all the other magnons. Therefore, the Bethe root $u_j$ and the momentum $p_j$  of the $j$th magnon are related by
\beq
u=\frac{1}{2}\cot\frac{p}{2} \la{up} \,.
\eeq
Having diagonalized $\hat T$ we have automatically diagonalized \textit{all} Hamiltonians obtained from this transfer matrix. For example, the Heisenberg Hamiltonian (\ref{Hxxx}) can be obtained as $\left.\frac{1}{2i}\frac{d}{d\lambda} \log \hat T(\lambda)\right|_{\lambda=0}$ so that the spectrum is simply
\beq
E=\frac{1}{2i} \left.\frac{d}{d\lambda} \log T(\lambda)\right|_{\lambda=0}=\sum_{j=1}^M \frac{1/2}{u_j^2+1/4}=\sum_{j=1}^M 2 \sin^2\frac{p_j}{2}\nn \,.
\eeq
By considering more derivatives of $\log \hat T$ at $\lambda=0$ we generate other local  Hamiltonians with longer range. To obtain the spectrum of these Hamiltonians we simple act with more derivatives on the logarithm of the eigenvalue (\ref{Teigen}). In particular, if we take more than  $L$ derivatives, wrapping interactions, where the range of the Hamiltonian is bigger than the size of the spin chain, will appear. 
Remarkably, for all such models the Bethe equations are just (\ref{BAEsu2}) since they diagonalize the transfer matrix!

A particularly interesting hamiltonian is
\beq
\hat H(g)= \frac{1}{4 i} \log \frac{ \hat T( g^2) }{\hat T(0)}+ h.c. \la{Htoy}
\eeq
If we think of $g^2$ as being an expansion parameter then we have an infinite range Hamiltonian where, at each order $g^{2n}$ in perturbation theory, the interaction range is $n$. In the notations of  \cite{Beisert:2004ry} we have
\beq
\hat H(g)%-\frac{L}{4i} \log\frac{i-g^2}{i+g^2} 
= \frac{g^2}{2} \sum_j \mathcal{H}_{j,j+1}+\frac{i  g^4}{4} \sum_j \[ \mathcal{H}_{j,j+1}, \mathcal{H}_{j+1,j+2}\]+ \dots  \la{Hper}
\eeq 
where $\mathcal{H}_{j,j+1}=1-\mathcal{P}_{j,j+1}$.  The spectrum of this Hamiltonian is then given by (\ref{Htoy}) where we simply replace the operators $\hat T(\cdot)$ by the corresponding eigenvalues $T(\cdot)$ to get
\beq
E=\frac{1}{4i} \sum_{j=1}^M \log\(   \frac{u_j+\frac{i}{2}-g^2}{u_j-\frac{i}{2}-g^2}\frac{u_j-\frac{i}{2}}{u_j+\frac{i}{2}}\)+\frac{1}{4i}\log\[1+\(\frac{g^2}{g^2+i}\)^{L} \prod_{j=1}^{M}\frac{u_j-\frac{3i}{2}-g^2}{u_j+\frac{i}{2}-g^2}\frac{u_j-\frac{i}{2}-g^2}{u_j+\frac{i}{2}-g^2} \]+c.c. \,. \la{spectrum}
\eeq
The first term comes from the first term in (\ref{Teigen}). It gives a contribution to the energy of the form $\sum \epsilon(p_j)$, that is a sum of the dispersion relations of $M$ individual magnons interacting through (\ref{BAEsu2}). The dispersion relation, when written in terms of $p$, yields 
\beq
\epsilon(p)= \frac{1}{2i} \log\(\frac{1-2g^2 e^{-ip/2}  \sin\frac{p}{2}}{1-2g^2 e^{+ip/2}  \sin\frac{p}{2}}\) = 2 g^2 \sin^2\frac{p}{2}+2 g^4 \sin p \sin^2\frac{p}{2}+\dots +  \frac{\(2 g^2\)^{n}}{n} \sin \frac{np}{2}\sin^n\frac{p}{2}+\dots \la{disp}
\eeq
The second term in (\ref{spectrum}), which comes from the second term in (\ref{Teigen}), is identically zero up to order $g^{2L}$, precisely when wrapping interactions appear! Thus, at order $g^{2L}$ the energy is given by a sum of dispersion relations of the form (\ref{disp}) plus this wrapping term, which entangles all $M$ magnons and is not writable as a sum of individual magnon energies. 
In terms of the magnon momentum we have
\beq
E=\sum_{j=1}^M \epsilon(p_j)+ g^{2L} \, \frac{1}{4i}\[i^{-L} \prod_{j=1}^M \(2 e^{-2ip_j}-e^{-ip_j}\)-c.c.\]+\mathcal{O}(g^{2L+1})\, \,. \la{energywithwrap}
\eeq
For example, for $1$ and $2$ magnons we get respectively 
\beq
E(p)= \epsilon(p)+ \frac{ g^{2L} }{2}\(\sin\(p+\frac{L\pi}{2}\)-2 \sin\(2p+\frac{L\pi}{2}\)\)+\mathcal{O}(g^{2L+2})\, \,, \la{1M}
\eeq
and
\beqa
\!\!\!\!\!\!\!\!E(p,k)\!\!&=&\!\! \epsilon(p)+\epsilon(k)+  g^{2L} \( s(p)+s(k)-\frac{s(0)}{2} -2s(p+k)\)
+\mathcal{O}(g^{2L+2})\, \,,\la{2M}
\eeqa
where $s(x)\equiv \sin(p+k+L\pi/2+x)$.
In particular, it is clear that the correction to the energy of the two magnon state is not of the form $\delta \epsilon(p)+\delta \epsilon(k)$. This was expected since at order $g^{2L}$ the interaction range covers the entire chain and the notion of asymptotic region where one can safely measure the dispersion relation of each magnon is destroyed \cite{Staudacher:2004tk,Bill}.

When we compare the type of corrections (\ref{1M}) and (\ref{2M}) we get from this long range model compared with those we got within the Hubbard model (\ref{prediction}) and (\ref{finalper2}) we see that while in the latter the coefficient of the order $g^{2L}$ wrapping correction to the energy
exhibited a strong $L$ dependence, from factors like $\cos^L(p/2)$, in the former this coefficient is $L$ independent and generically of order $1$. 
If this feature is generic then the weak or strong $L$ dependence of the $g^{2L}$ prefactors to the corrections of $2$ magnons states in $\mathcal{N}=4$ super Yang-Mills could be a sign, respectively, of a \textit{fundamental} description in terms of a long ranged exact integrable Hamiltonian or of a local description in terms of an hidden level of particles in the Beisert-Staudacher nested Bethe ansatz equations. 
In this context, the perturbative computation of \cite{w1} and \cite{w2}, if generalized to a more general setup with $2$ magnons and generic $L$ could provide very useful hints about the nature of the fundamental integrable structure of $\mathcal{N}=4$ super Yang-Mills.

\subsection{Generalizations and transcendentality} \la{Toy2gen}
In the previous section we saw that we could easily generate (long-ranged) integrable Hamiltonians by considering
\beq
\hat H= \left.\frac{1}{4i}\sum_{n=1}^{\infty} \, a_n\,\frac{g^{2n} }{n!} \frac{d^n}{d\lambda^n} \log \hat T(\lambda) \right|_{\lambda=0} +h.c.  \la{Hgen}
\eeq
The spectrum of such Hamiltonians is immediately given by this expression with the operator $\hat T(\lambda)$ replaced by the corresponding eigenvalue (\ref{Teigen}). At order $g^{2n}$ these Hamiltonians are local with interactions of range $n$. If we truncate the expansion at a given order $m$ by setting $a_{n>m}=0$, then for chains of length larger than $m$ we have no wrapping interactions and the energy is simply given by a sum of dispersion relations $\sum_j \epsilon(p_j)$,  with\footnote{If the $a_n$ are not real -- which from the definition \ref{Hgen} is a perfectly reasonable possibility -- we should take the real part of this expression.} 
\beq
\epsilon(u)=\frac{1}{2i}\sum_{n=1}^{\infty} \frac{a_n\,g^{2n}}{n} \( \frac{1}{(u-i/2)^n}-\frac{1}{(u+i/2)^n}\) \la{dispersion}
\eeq
If, on the other hand, we consider an infinie sum or, alternatively, if we truncate the expansion in $g^2$ at an order $m>L$,  the spectrum (starting at wrapping order $g^{2L}$) will no longer be a sum of individual dispersion relations. In particular, precisely at order  $g^{2L}$ we obtain
\beq
E=\sum_{j=1}^M\epsilon(u_j)+\frac{g^{2L}}{4i}\[\frac{a_L}{i^L}e^{-i P} \prod_{j=1}^{M}\frac{u_j-3i/2}{u_j+ i/2}   -c.c.\]+\mathcal{O}(g^{2L+2}) \la{Hleadin}
\eeq
where the second term takes into account the wrapping interactions. It is interesting to notice that for all these families of long ranged Hamiltonians the expression for the wrapping interactions is quite simple and absolutely universal. The example (\ref{Htoy}) we considered corresponds to $a_n=1$ but as we see any choice of $a_n$ will lead to a solvable problem. A particularly funny example would be 
\beq
\hat H= \left.\frac{1}{2i}\sum_{n=1}^{\infty} \, C_n\,\frac{g^{2n} }{(2n-1)!} \frac{d^{2n}}{d\lambda^{2n}} \log \hat T(\lambda) \right|_{\lambda=0} +h.c.  \,,  \nn
\eeq
with $C_n$ the Catalan numbers, for which we find the curious expression
$$
\epsilon(u)=\frac{g}{i}\( \frac{1}{X^+(u)}-\frac{1}{X^-(u)} \)\,\,, \,\, X^\pm(u)\equiv \frac{u\pm \frac{i}{2}+\sqrt{\(u\pm \frac{i}{2}\)^2-4g^2}}{2g}
$$
for the dispersion relation. As a function of the Bethe roots this is precisely the dispersion relation appearing in the BDS equations \cite{Beisert:2004hm} and even in the full AdS/CFT Bethe equations \cite{BS}. However, unfortunately, as a function of the magnon momenta this is not the same as (\ref{Eninf}) because the relation between $u$ and $p$ in our toy models is always of the form (\ref{up}) instead of (\ref{uuinf}). In other words, although the desired dispersion relation can easily be obtained, the Bethe roots are always quantized via the usual Heisenberg spin chain Bethe equations.

We can generalize this construction of long ranged Hamiltonians for other symmetry groups as well. For example, for $SU(N)$ the transfer matrix in the fundamental representation takes exactly the same form (\ref{Tsu2}) as for the $SU(2)$ chains except that the operators being multiplied now live in $h_j\otimes h_0$ where both $h_0$ and $h_j$ are copies of $\mathbb{C}^N$. Thus, we can still consider Hamiltonians of the form (\ref{Htoy}) with a $g^2$ perturbative expansion (\ref{Hper}). As before, they will  be long ranged Hamiltonians  where the Hamiltonian range at  order $g^{2n}$ is $n$. The only thing that changes is the expression for the eigenvalue (\ref{Teigen}). For an $SU(N)$ spin chain we have $N-1$ types of roots and $K_n$ Bethe roots of the type $n=1,\dots,N-1$. The transfer matrix eigenvalue reads\footnote{The $SU(N)$ Bethe equations can be immediatly obtained by canceling the apparent $\lambda$ poles in this expression.}
\beqa
T_{SU(N)}(\lambda)= \prod_{j=1}^{K_1}\frac{\lambda-u_j^{(1)}-i/2}{\lambda-u_j^{(1)}+i/2}
+\(\frac{\lambda}{i+ \lambda}\)^L   \sum_{n=1}^{N-1} \prod_{j=1}^{K_{n}}\frac{\lambda-u_j^{(n)}+\frac{n+2}{2}i}{\lambda-u_j^{(n)}+\frac{n}{2}i} \prod_{j=1}^{K_{n+1}}\frac{\lambda-u_j^{(n+1)}+\frac{n-1}{2}i}{\lambda-u_j^{(n+1)}+\frac{n+1}{2}i} \nn
\eeqa
We see again that when we compute the first few local charges expanding $\log T(u)$ around $u=0$ only the first term gives a non-vanishing contribution. As before we can consider Hamiltonians of the form (\ref{Hgen}). The dispersion relation for the $SU(N)$ magnons is exactly the same as for $SU(2)$ while the leading wrapping correction to the spectrum can be computed as before to yield the generalization of (\ref{Hleadin}),
\beq
E=\sum_{j=1}^{K_1}\epsilon(u_j^{(1)})+\frac{g^{2L}}{4i}\[\frac{a_L}{i^L}\, e^{-iP}\sum_{n=1}^{N-1} \prod_{j=1}^{K_{n}}\frac{u_j^{(n)}-\frac{n+2}{2}i}{u_j^{(n)}-\frac{n}{2}i} \prod_{j=1}^{K_{n+1}}\frac{u_j^{(n+1)}-\frac{n-1}{2}i}{u_j^{(n+1)}-\frac{n+1}{2}i}  -c.c.\]+\mathcal{O}(g^{2L+2}) \nn \,.
\eeq
where again $P=\frac{1}{i}\sum_j \log\frac{u_j^{(1)}+i/2}{u_j^{(1)}-i/2}$ is the state total momentum.

All these considerations can be trivially generalized both to non-compact spin chains and to supersymmetric ones. For example, for the $SL(2)$ spin chain we have \cite{Kulish:1981gi,Faddeev:1996iy}
\beq
T_{sl(2)}(\lambda)=\prod_{j=1}^M \frac{\lambda-u_j-i/2}{\lambda-u_j +i/2}+\sum_{n=1}^{\infty}
\(\frac{\lambda}{in+ \lambda}\)^L    \prod_{j=1}^{M}\frac{\(\lambda-u_j-i/2\)^2}{\(\lambda-u_j+\frac{2n-1}{2}i\)\(\lambda-u_j+\frac{2n+1}{2}i\)} \la{Tsl2}
\eeq
with the Bethe equations, following from canceling the poles in this expression, reading
\beq
\(\frac{u_j+i/2}{u_j-i/2}\)^L=\prod_{k\neq j}^{M}\frac{u_j-u_k-i}{u_j-u_k+i} \, \la{BAEsl2}
\eeq
which differ from (\ref{BAEsu2}) by a simple sign in the r.h.s. Again, by expanding the log of the transfer matrix around $\lambda=0$ we see that only the first term contributes until the $L$'th derivative is taken. Thus if we consider an Hamiltonian of the form (\ref{Hgen}) we will have, up to order $g^{2L}$, the energy as a sum of the same dispersion relations (\ref{dispersion}). In particular if the constants $a_n$ are algebraic numbers then so will be $\sum_{j} \epsilon(u_j)$ when truncated to order $g^{2L}$ because clearly  the solutions to (\ref{BAEsl2}) are also algebraic (complex) numbers. 

However, precisely at order $g^{2L}$ the second term in (\ref{Tsl2}) starts contributing and we find 
\beq
E=\sum_{j=1}^{M} \epsilon(u_j)+\frac{g^{2L}  }{4i}\[\frac{a_{L}}{i^L}e^{-i P} \,\sum_{n=1}^{\infty}
\frac{1}{n^L}    \prod_{j=1}^{M}\frac{\(u_j+i/2\)^2}{\(u_j-\frac{2n-1}{2}i\)\(u_j-\frac{2n+1}{2}i\)}-c.c.\] \la{wrapsl2} \,.
\eeq
This new wrapping term differs from the one computed for the compact groups $SU(N)$ by the fact that it is given by an infinite sum of terms. Thus even if $u_j$ and $a_n$ are perfectly algebraic numbers the energy of this state will only be algebraic up to order $g^{2L}$, when this infinite sum will give a transcendental contribution!

Let us consider a few examples. We chose $a_L=1$ for simplicity.
For $L=4$ the one magnon state with momentum $2\pi/4$ will be corrected to
\beq
E=\epsilon(p)-g^{6}\(1-\zeta(3) \) +\mathcal{O}(g^{8}) \nn
\eeq
while for example for a two magnon state with $L=5$ and momenta\footnote{For two magnon with opposite momenta the $sl(2)$ equations are trivially solved exactly as explained in the end of section \ref{BDSper} for the $su(2)$ chain. For the non-compact chain the effect of the second magnon is simply to renormalize $L\to L+1$ instead of $L\to L-1$ as we had for the $SU(2)$ chain. Thus we obtain $p_1=-p_2=\frac{2\pi n}{L+1}$ for the two magnon state with opposite symmetric momenta.} $p_1=-p_2=p=2\pi/6$ we get
\beq
E=\epsilon(p)+\epsilon(-p)+\frac{g^{10}}{4}\(1-2\zeta(3)+2\zeta(5)\)+\mathcal{O}(g^{12}) \nn
\eeq
In (\ref{table}) in the introduction we listed a couple of additional examples. 

Although, this model is not immediately related to the (non-compact sector of) AdS/CFT Bethe equations which are much more complicated than (\ref{BAEsl2}), it is still interesting to see that transcendentality very naturally appears due to the non-compact nature of the transfer matrix. In particular, if an extra level of hidden degrees of freedom is to be discovered then the transcendental numbers present in the dressing factor could be an important hint. A more fundamental $PSU(2,2|4)$ symmetric transfer matrix in the field strength representation would also be given by an infinite  sum of terms since the representation is infinite dimensional. It would be spectacular if a relatively simple extended transfer matrix with some extra degrees of freedom included and only simple algebraic expressions could lead to the intricate structure of the full dressing factor where transcendental numbers abound. Probably the correct place to try to reverse engineer and find this extra level of hidden particles is the transfer matrix rather than the Bethe equations.

\section*{Acknowledgments}
We would like to thank N.~Gromov, V.~Kazakov, A.~Kozak, S. Schafer-Nameki, J.~Polchinski, P.~Ribeiro,  and K.~Zarembo for several interesting discussions and insightful comments. JP is funded by the FCT fellowship SFRH/BPD/34052/2006.  PV is funded by the Funda\c{c}\~ao para a Ci\^encia e Tecnologia fellowship {SFRH/BD/17959/2004/0WA9}. PV would like to thank KITP and its several members for the warm hospitality and for partially funding this work. PV would like to thank Caltech Institute of Technology for warm hospitality.  This research was supported in part by the National Science Foundation under Grant No. NSF PHY05-51164.

\setcounter{section}{0}
\setcounter{subsection}{0}

%%%%%%%%%%%%%%%%%%%%%%%%%%%%%%%%%%%%%
%%%%%%%%%%%%%%%%%%%%%%%%%%%%%%%%%%%%%
%%%%%%%%%%%%%%%%%%%%%%%%%%%%%%%%%%%%%

\appendix{Wave function of the Hubbard magnon}
\label{AppSch}

In this section we shall study states made out of two fundamental particles $o$ and $\updownarrow$, 
\beqa
|\Psi\rangle=\sum_{n,n'}\psi(n,n') \(a^{\dagger}_{n,\updownarrow} a^\dagger_{n',o} \)  |\uparrow \dots\uparrow \rangle \,.
\nn
\eeqa
Acting on this state with the Hamiltonian (\ref{HHub}) we can find the form of the wave function $\psi(n,n')$ so that the state is an eigenstate,
$$
H|\Psi\rangle= E |\Psi\rangle \,.
$$
This gives the equation 
\beqa
E\,\psi(n,n') \!\!\!&=&\!\!\!
-t\[e^{i\phi_{\updownarrow}} \psi(n+1,n')+e^{-i\phi_{\updownarrow}} \psi(n-1,n')+e^{i\phi_{o}} \psi(n,n'+1)+e^{-i\phi_o} \psi(n,n'-1)\] \nn \\
(E+U)\,\psi(n,n)\!\!\!&=&\!\!\! 
-t \[e^{i\phi_{\updownarrow}} \psi(n+1,n)+e^{-i\phi_{\updownarrow}} \psi(n-1,n)+e^{i\phi_{o}} \psi(n,n+1)+e^{-i\phi_o} \psi(n,n-1)\] \nn
\eeqa
with $n' \neq n$. A plane wave superposition 
\beqa
\psi(n,n')&=&e^{-i \phi_\updownarrow n-i\phi_o n'}
\(A e^{iqn+iq'n'}+B e^{iq'n+iqn' } \)\ ,\ \ \ \ \ \ \ \ n<n' \la{magnonwfL}\\
\psi(n,n')&=&e^{-i \phi_\updownarrow n-i\phi_o n'}
\(C e^{iqn+iq'n'}+D e^{iq'n+iqn' } \)\ ,\ \ \ \ \ \ \ \ n>n' \la{magnonwfR}
\eeqa
solves the first equation and yields 
\beq
E=-2\,t\,\cos(q )-2\,t\,\cos(q' )  \,. \nn \la{energy}
\eeq
Continuity of the wave function at $n=n'$ gives 
$$
A+B=C+D
$$
and the second equation reduces to 
%{\tiny
%$$
%(A+B)\(e^{iq}+e^{-iq}+e^{iq'}+e^{-iq'}+\frac{1}{g}\)=
%(C+B)\(e^{iq}+e^{-iq'}\)+(A+D)\(e^{-iq}+e^{iq'}\)
%$$
%$$
%B\(e^{-iq}+e^{iq'}-\frac{1}{g}\)+A\(e^{iq}+e^{-iq'}+\frac{1}{g}\)
%=C\(e^{iq}+e^{-iq'}\)+D\(e^{-iq}+e^{iq'}\)
%$$
%$$
% A \( 2ig \sin(q)-2ig\sin(q')+1 \)
%=-B+C\( 2ig \sin(q)-2ig\sin(q')\)
%$$}
%$$
% \frac{A}{C}= \frac{2g \sin(q)-2g\sin(q') +i\frac{B}{C}}{ 2g \sin(q)-2g\sin(q')-i }
%$$
\beq
 A\(2g \sin(q)-2g\sin(q')-i \)= C\(2g \sin(q)-2g\sin(q') \)+iB \,. \la{genSmatrix}
\eeq
In an infinitely large volume we might look for bound states with exponentially decaying wave functions. These are only possible if two of the exponentials in (\ref{magnonwfL}) and (\ref{magnonwfR}) disappear. Parametrizing the momenta as $q=p/2-i\beta$ and $q'=p/2+i \beta$ we see that we need $B=C=0$ and $A=D\neq 0$. Then equation (\ref{genSmatrix}) fixes
\beq
\sinh\beta=\sinh\beta_\infty \equiv -\frac{1}{4g\cos\frac{p}{2}}  \la{betainf2}
\eeq
in which case
%$$
%\psi(n,n')= e^{i\frac{p}{2} (n+n') } e^{-\beta_{\infty} |n-n'| }
%$$
$$
\psi(n,n')=e^{-i \phi_\updownarrow n-i\phi_o n'}e^{i\frac{p}{2} (n+n') } e^{-\beta |n-n'| }
$$
and the energy (\ref{energy}) reads
\beq
E=\epsilon_\infty(p)\equiv -4t\cos\frac{p}{2}\cosh\beta_\infty=- U\sqrt{1+16g^2\cos^2\frac{p}{2}} \,. \nn
\eeq
Comparing with the energy (\ref{energy}) of two particles with momentum $p/2$ we conclude that the magnon has a relative binding energy
$$
\cosh \beta -1\,.
$$

On the other hand, at finite volume we impose periodicity 
$$
\psi(n+L,n')=\psi(n,n')\ , \ \ \ \ n<n'<n+L
$$
and
$$
\psi(n,n'+L)=\psi(n,n')\ , \ \ \ \ n'<n<n'+L
$$
of the wave function  to obtain
$$
Ce^{i(q-\phi_\updownarrow)L}=A\ ,\ \ \ \ \ \ De^{i(q'-\phi_\updownarrow)L}=B
$$
and
$$
Ae^{i(q'-\phi_o)L}=C\ ,\ \ \ \ \ \ Be^{i(q-\phi_o)L}=D
$$
This immediately gives total momentum quantization
$$
e^{i(q+q'-\phi_\updownarrow-\phi_o)L}=1
$$
and the relation  
$$
\frac{B}{C}=\frac{1-e^{i(q-\phi_\updownarrow) L}}{1-e^{i(q-\phi_o) L}}
$$
From (\ref{genSmatrix}) we obtain 
$$
e^{i(q-\phi_\updownarrow)L}= \frac{2g \sin(q)-2g\sin(q') +i\frac{1-e^{i(q-\phi_\updownarrow) L}}{1-e^{i(q-\phi_o) L}} }
{ 2g \sin(q)-2g\sin(q')-i}
$$
which can be rewritten as
$$
e^{i(q-\phi_o)L}= \frac{g \(\sin(q)-\sin(q')\) \(1+e^{i(\phi_\updownarrow-\phi_o)L} \)+
\sqrt{g^2 \(\sin(q)-\sin(q')\)^2 \(1-e^{i(\phi_\updownarrow-\phi_o)L} \)^2  -e^{i(\phi_\updownarrow-\phi_o)L} }}
{ 2g \sin(q)-2g\sin(q')-i}
$$
This is precisely what one obtains from the Lieb-Wu equations (\ref{HBAE}) for the Bethe roots ${q,q',u}$ eliminating the auxiliar variable $u$ using its Bethe equation. Moreover, in the case of identical twists this becomes the simple S-matrix
$$
e^{i(q-\phi_o)L}= \frac{2g \sin(q)-2g\sin(q') +i}
{ 2g \sin(q)-2g\sin(q')-i}
$$
If again we look for solutions in the form $q=p/2-i\beta$ and $q'=p/2+i \beta$ so that
 \beq
E= \epsilon(p)=-4t\cos\frac{p}{2}\cosh\beta \,, \nn
\eeq
we obtain
\beq
e^{i (p-\phi_o-\phi_\updownarrow) L}=1 \la{momquant}
\eeq
and 
\beq
4g \cos\frac{p}{2}\sinh\beta=-\frac{\sinh (\beta L)}{\cosh( \beta L)-\cos L\( \frac{p}{2} -\phi_o\)}\,. \nn
\eeq
Notice that the result is symmetric under the exchange of twists since (\ref{momquant}) implies
$$
\cos L\( \frac{p}{2} -\phi_o\)=\cos L\( \frac{p}{2} -\phi_\updownarrow\)=\cos \frac{L}{2}\( \phi_\updownarrow-\phi_o\) \,.
$$
In the case of equal twists the expression reduces to
\beq
4g \cos\frac{p}{2}\sinh\beta=-\frac{\sinh (\beta L)}{\cosh( \beta L)\pm 1}\,, \nn
\eeq
and for BDS twists obeying $e^{iL(\phi_o-\phi_\updownarrow)}=-1$ gives
\beq
4g \cos\frac{p}{2}\sinh\beta=-\tanh\(\beta L\) \,. \nn
\eeq
In any case, for large $L$ the r.h.s. can be replaced by $-1$ and we recover the infinite volume result (\ref{betainf2}). 

\appendix{Dressed BDS BAE with twists}
\label{AppBAE}
This appendix complements the computations and results of sections (\ref{corrected}) and (\ref{BDSper}). As we see from (\ref{BDSupgrade}), (\ref{deltaE}) and (\ref{deltabeta}) a crucial quantity we need to compute the leading wrapping corrections is $\Delta_n$ defined in (\ref{uD}).
To compute $\Delta_n$ we focus on the equation for a single constituent of the Takahashi triplet, say on  equation (\ref{bae1}) for $q_n$. Expanding this equation using (\ref{uD}) and (\ref{uper}) 
yields 
\beqa
\Delta_n\simeq \Delta_n^{(0)}\equiv i e^{i\(\frac{p_n}{2}+i\beta_n -\phi_o \)L}\prod_{m\neq n}^{M} 
\frac{u^\infty_m-u^\infty_m+i}{u^\infty_m-u^\infty_m} =\mathcal{O} (e^{-\beta_n L}) \,,
\eeqa
to leading order. However, as seen from the above mentioned equations (\ref{BDSupgrade}), (\ref{deltaE}) and (\ref{deltabeta})  what we really need is the imaginary part of $\Delta_n$.  Due to the corrected BDS equations (\ref{BDSupgrade}) the imaginary part of $\Delta_n^{(0)}$ is of order $e^{- \beta_n L}$ for generic twists and of order $e^{-3 \beta_n L}$ when (\ref{gal}). Indeed , the imaginary part of $\Delta_n^{(0)}$ can be simplified using the BDS corrected equations (\ref{BDSupgrade}) to give
\beqa
\Im\,( \Delta_n^{(0)})\simeq -i \Delta_n^{(0)} \,e^{i\(\phi_o-\phi_\updownarrow\)\frac{L}{2}} 
\, \cos\frac{L}{2}(\phi_o-\phi_\updownarrow)\,  
\eeqa 
which indeed vanishes when (\ref{gal}). Thus, in this case we need to expand the equations for $q_n$ further, 
\beqa
\Delta_n\simeq \Delta_n^{(0)} -i  \(\(\Delta_n^{(0)}\)^2+ \sum_{m\neq n}^{M} \frac{\Delta_n^{(0)}\Delta_m^{(0)}}{(u^\infty_m-u^\infty_n)(u^\infty_m-u^\infty_n+i)} \)  \la{Deltaspecial} \,.
\eeqa 
To find the corrected BDS equations we multiply the equations for the Takahashi triplet to get
\beqa
e^{i (p_n-\phi_\updownarrow-\phi_o)L} \prod_{m\neq n} \frac{u^\infty_n -u^\infty_m-i}{u^\infty_n -u^\infty_m+i}=  \prod_{m=1}^M \frac{u^\infty_n-u^\infty_m-i}{u^\infty_n -u^\infty_m +i}\frac{u_n-u_m-i}{u_n-u_m+i} \frac{r_{m,n}}{r_{n,m}}\frac{r_{m,n+M}}{r_{n,m+M}} \la{triplet}
\eeqa
where 
\beqa
r_{n,m}\equiv \frac{u_n-2 g \sin q_m-i/2}{u_n-2 g \sin q_m +i/2}  \,.
\eeqa
Notice that so far no approximation whatsoever was done and (\ref{triplet}) is an exact relation. Expanding now the r.h.s. of (\ref{triplet}) using (\ref{uD}) and (\ref{uper})
we find  the corrected (twisted) BDS equations (\ref{BDSupgrade})

Again, when the twists are chosen as in (\ref{gal}) we need to be more careful and expand our expressions further. By expanding the product of the equations for the Takahashi triplet to order $e^{-2\beta L}$ we get the same (\ref{BDSupgrade}) with
\beqa
\phi_{nm}=-\frac{\Im(\Delta_n)}{(u_n-u_m)^2+1}\(\frac{1}{u_n} \tan^2\frac{p_n}{2}+\frac{2}{u_n-u_m}\)+\frac{2(u_n-u_m)\Delta_n\Delta_m}{((u_n-u_m)^2+1)^2}- (n\leftrightarrow m) \la{phispecial}
\eeqa
instead of (\ref{dressing}).

Finally, in (\ref{deltaE}) the energy is corrected because the momenta change (the second term) and because the functional dependence on the momenta changes (the first term). For completeness let us mention why the correction due to the change in functional dependence is so simple. This is simply because 
\beqa
-4t \cos\frac{p_n}{2}\cosh \(\beta_\infty(p_n)+\delta\beta_n\)=  -U \(-4g \cos\frac{p_n^\infty}{2}\sinh \beta_\infty(p_n^\infty) \)\delta \beta_n=-U \delta\beta_n
\eeqa
since $-4g \cos\frac{p_n^\infty}{2}\sinh \beta_\infty(p_n^\infty) =1$.
\appendix{Luscher with magnon-magnon S-matrix at strong coupling}
In this appendix we expand the results of \cite{Janik:2007wt}\footnote{The notation in this paper differs from ours by $p_{there}=p_{here}+\pi$, $g_{there}=\sqrt{2} g_{here}$. Moreover, we have an extra factor of $U$ multiplying the magnon dispersion relation. This shift in $p$ introduces an  annoying  $(-1)^L$ factor. In order to avoid carrying it around we consider $L$ even in this appendix. } concerning the BDS Luscher term to the next order in $1/g$.  
The computation in question is the $\mu$-term
\begin{eqnarray*}
\delta\epsilon(p) & = & \frac{1}{2} Res_{k=k^*}e^{iL k} \left[\epsilon^{\prime}(p)-\epsilon^{\prime}(k)\right]\frac{u(p)-u(k)+i}{u(p)-u(k)-i} \,\,+\,\,c.c.\\
 & = &  e^{iL k^* } \left[\epsilon^{\prime}(p)-\epsilon^{\prime}(k^*)\right] \frac{i}{-u^{\prime}(k^*)}   \,\,+\,\,c.c. \\
\end{eqnarray*}
where  
\beq
u(p)=\frac{1}{2}\tan\frac{p}{2}\sqrt{1+16g^2 \cos^2\frac{p}{2}} \nn
\eeq
and $k^*$ is defined by the pole condition 
\beq
u(p)-u(k^*)+i=0 \,. \nn
\eeq
For large $g$, this gives
\beq
k^*=\pi +\frac{i}{2g}\sec\frac{p}{2} +\frac{1}{4g^2}\tan\frac{p}{2} \sec^2\frac{p}{2}
+\frac{i}{384 g^3}  \(23 \cos p -49\)   \sec^5\frac{p}{2} +
\dots \nn
\eeq
which yields 
\beq
\delta\epsilon(p)= \Re \(-\frac{U}{g} \sec \frac{p}{2} +  \frac{iU}{g^2}\tan\frac{p}{2} \sec^2\frac{p}{2} 
%-\frac{1}{128g^ 3} \( 23 \cos p -49\) \sec^5\frac{p}{2}
+\dots \) 
\exp -\frac{L}{2g}\sec\frac{p}{2}\(1 -\frac{i}{2g}\tan\frac{p}{2} \sec\frac{p}{2} +\dots \) \nn
\eeq

\appendix{Many particle Luscher formula and $SO(\infty)$ sigma model}
In this appendix we apply our conjectured expression (\ref{conjecture}) to the study of finite size corrections to the energy of many particle states in the $SO(\infty)$ sigma model. The motivation for such computation is twofold. 
On the one hand, it is a very interesting model where the single particle corrections are very well understood \cite{Luscher:1982uv,Luscher:1985dn,Klassen:1990ub}. On the other hand, large $N$ sigma models often allow for the computations to be done in a vast panoply of ways and in particular it might be simple to cross check our conjecture through some saddle point computations. 

To apply (\ref{conjecture}) we need to compute 
$S_{a_1, b_1}^{a_1,b_2}(p_1,q)S_{a_2, b_2}^{a_1,b_3}(p_2,q)\dots S_{a_M, b_M}^{a_M,b_1}(p_2,q)$ 
which is precisely the transfer matrix $T(\theta;\{\theta_j\})={\rm tr}_0 \(\hat S_{1,0}(\theta_1-\theta)\dots \hat S_{M,0}(\theta_M-\theta)\)$ if we parametrize $p_j=m\sinh\theta_j$ and $q=m\sinh\theta$. In the large $N$ limit the $S$-matrix tends to unity but since we will sum over a large number $N$ of such terms, the sum we need to compute converges to a nice quantity, namely, using the normalizations of \cite{Klassen:1990ub}, 
\beq
S_{a_1, b_1}^{a_1,b_2}(p_1,q)S_{a_2, b_2}^{a_1,b_3}(p_2,q)\dots S_{a_M, b_M}^{a_M,b_1}(p_2,q) - \delta_{b_1}^{b_2} \dots \delta_{b_M}^{b_1}  \rightarrow \sum_{j=1}^M \frac{2\pi i}{\sinh(\theta-\theta_j)} \nn
\eeq
where we sum over the index of the virtual fundamental particle $b_j$.
There are no bound states in this theory and therefore the $\mu$-term is zero.
Thus, we will consider only $F$-term like contributions. Finally we have $dq\(\epsilon'(q)-\epsilon'(p_j)\)=m d\theta  \sinh(\theta-\theta_j)/\sinh\theta_j$ and $e^{iqL}=e^{imL\sinh\theta}$.  Will all ingredients at hand it is then straightforward to apply (\ref{conjecture}) to find 
\beq
\delta E(\theta_1)= \frac{\delta m}{\cosh\theta_1}  \la{onept}
\eeq
for a single particle state and 
%\beq
%\delta E(\theta_1,\theta_2)= \frac{\delta m}{\cosh\theta_1}+\frac{\delta m}{\cosh\theta_2}+\delta \epsilon_{int}(\theta_1,\theta_2) \la{twopt}
%\eeq
\beq
\delta E(\theta_1,\dots,\theta_M)=\sum_{j=1}^{M} \frac{\delta m}{\cosh\theta_j}+ \sum_{i<j}^M \delta \epsilon^{ij}_{int}(\theta_1,\theta_2) \la{twopt}
\eeq
for a many particle state. Here 
\beq
\frac{\delta m}{m}=\int d\theta \,e^{-mL \cosh\theta}= 2 K_0(mL)\,, \nn
\eeq
in agreement with \cite{Luscher:1982uv}, and 
%\beq
%\delta \epsilon_{int}=\int d\theta e^{-mL \cosh\theta} \(\frac{\cosh(\theta-\theta_1)}{\cosh(\theta_1)\cosh(\theta-\theta_2)}+\frac{\cosh(\theta-\theta_2)}{\cosh(\theta_2)\cosh(\theta-\theta_1)}\) \la{interaction}
%\eeq
\beq
\delta \epsilon_{int}^{ij}=m\int d\theta \,e^{-mL \cosh\theta} \(\frac{\cosh(\theta-\theta_i)}{\cosh(\theta_i)\cosh(\theta-\theta_j)}+\frac{\cosh(\theta-\theta_j)}{\cosh(\theta_j)\cosh(\theta-\theta_i)}\) \la{interaction}
\eeq
These expressions have a clear physical meaning. As for (\ref{onept}) we get the expected dependence on the momenta since we know that for a single excitation the form of the dispersion relation will change solely due to the change in the particle mass when the system is put in finite volume and thus we expect $\delta \epsilon=\delta \sqrt{m^2+p^2}=m \delta m/\sqrt{m^2+p^2}=\delta m/\cosh(\theta)$ precisely as found. For the many particle particle state we get expression (\ref{twopt}) where we see that the energy of the state is corrected due to the mass shift of each particle and by an extra interaction term (\ref{interaction}). Probably this interaction correction to the energy of the many particle state  could be independently determined through some sort of saddle point computation. This would be extremely interesting as it could help to partially confirm (or correct) (\ref{conjecture}).

%%%%%%%%%%%%%%%%%%%%%%%%%%%%%%%%%%%%%
%%%%%%%%%%%%%%%%%%%%%%%%%%%%%%%%%%%%%
%%%%%%%%%%%%%%%%%%%%%%%%%%%%%%%%%%%%%

%%%%%%%%%%%

\bibliographystyle{JHEP}
\renewcommand{\refname}{Bibliography}
\addcontentsline{toc}{section}{Bibliography}
\bibliography{GMbib}

\begin{thebibliography}{99}
\small
%\cite{Minahan:2002ve}
\bibitem{MZ}
  J.~A.~Minahan and K.~Zarembo,
  ``The Bethe-ansatz for N = 4 super Yang-Mills,''
  JHEP {\bf 0303}, 013 (2003)
  [arXiv:hep-th/0212208].
  %%CITATION = JHEPA,0303,013;%%

%\cite{Beisert:2003jj}{Beisert:2003yb}
\bibitem{Beisert:2003jj}
  N.~Beisert,
   ``The complete one-loop dilatation operator of N = 4 super Yang-Mills
  theory,''
  Nucl.\ Phys.\  B {\bf 676}, 3 (2004)
  [arXiv:hep-th/0307015].
  %%CITATION = NUPHA,B676,3;%%

%\cite{Beisert:2003yb}
\bibitem{Beisert:2003yb}
  N.~Beisert and M.~Staudacher,
  ``The N = 4 SYM integrable super spin chain,''
  Nucl.\ Phys.\  B {\bf 670} (2003) 439
  [arXiv:hep-th/0307042].
  %%CITATION = NUPHA,B670,439;%%

%\cite{Beisert:2004hm}
\bibitem{Beisert:2004hm}
  N.~Beisert, V.~Dippel and M.~Staudacher,
  ``A novel long range spin chain and planar N = 4 super Yang-Mills,''
  JHEP {\bf 0407} (2004) 075
  [arXiv:hep-th/0405001].
  %%CITATION = JHEPA,0407,075;%%

%\cite{Kazakov:2004qf}
\bibitem{Kazakov:2004qf}
  V.~A.~Kazakov, A.~Marshakov, J.~A.~Minahan and K.~Zarembo,
  ``Classical / quantum integrability in AdS/CFT,''
  JHEP {\bf 0405}, 024 (2004)
  [arXiv:hep-th/0402207].
  %%CITATION = JHEPA,0405,024;%%

%\cite{Beisert:2005bm}
\bibitem{Beisert:2005bm}
  N.~Beisert, V.~A.~Kazakov, K.~Sakai and K.~Zarembo,
  ``The algebraic curve of classical superstrings on AdS(5) x S**5,''
  Commun.\ Math.\ Phys.\  {\bf 263}, 659 (2006)
  [arXiv:hep-th/0502226].
  %%CITATION = CMPHA,263,659;%%

%\cite{Arutyunov:2004vx}
\bibitem{Arutyunov:2004vx}
  G.~Arutyunov, S.~Frolov and M.~Staudacher,
  ``Bethe ansatz for quantum strings,''
  JHEP {\bf 0410}, 016 (2004)
  [arXiv:hep-th/0406256].
  %%CITATION = JHEPA,0410,016;%%

\bibitem{BS}
  N.~Beisert and M.~Staudacher,
  ``Long-range PSU(2,2|4) Bethe ansaetze for gauge theory and strings,''
  Nucl.\ Phys.\  B {\bf 727} (2005) 1
  [arXiv:hep-th/0504190].
  %%CITATION = NUPHA,B727,1;%%

%\cite{Staudacher:2004tk}
\bibitem{Staudacher:2004tk}
  M.~Staudacher,
  ``The factorized S-matrix of CFT/AdS,''
  JHEP {\bf 0505}, 054 (2005)
  [arXiv:hep-th/0412188].
  %%CITATION = JHEPA,0505,054;%%

%\cite{Beisert:2005tm}
\bibitem{Beisert:2005tm}
  N.~Beisert,
  ``The su(2|2) dynamic S-matrix,''
  arXiv:hep-th/0511082.
  %%CITATION = HEP-TH/0511082;%%

%\cite{Beisert:2006qh}
\bibitem{Beisert:2006qh}
  N.~Beisert,
  ``The Analytic Bethe Ansatz for a Chain with Centrally Extended su(2|2)
  Symmetry,''
  J.\ Stat.\ Mech.\  {\bf 0701} (2007) P017
  [arXiv:nlin/0610017].
  %%CITATION = JSTAT,0701,P017;%%
  
  %\cite{Arutyunov:2006yd}
\bibitem{Arutyunov:2006yd}
  G.~Arutyunov, S.~Frolov and M.~Zamaklar,
  ``The Zamolodchikov-Faddeev algebra for AdS(5) x S**5 superstring,''
  JHEP {\bf 0704}, 002 (2007)
  [arXiv:hep-th/0612229].
  %%CITATION = JHEPA,0704,002;%%

%\cite{Beisert:2006ez}
\bibitem{Beisert:2006ez}
  N.~Beisert, B.~Eden and M.~Staudacher,
  ``Transcendentality and crossing,''
  J.\ Stat.\ Mech.\  {\bf 0701}, P021 (2007)
  [arXiv:hep-th/0610251].
  %%CITATION = JSTAT,0701,P021;%%

%\cite{Beisert:2006ib}
\bibitem{Beisert:2006ib}
  N.~Beisert, R.~Hernandez and E.~Lopez,
  ``A crossing-symmetric phase for AdS(5) x S**5 strings,''
  JHEP {\bf 0611}, 070 (2006)
  [arXiv:hep-th/0609044].
  %%CITATION = JHEPA,0611,070;%%
 
 %\cite{Arutyunov:2006ak}
\bibitem{Arutyunov:2006ak}
  G.~Arutyunov, S.~Frolov, J.~Plefka and M.~Zamaklar,
  ``The off-shell symmetry algebra of the light-cone AdS(5) x S**5
  superstring,''
  J.\ Phys.\ A  {\bf 40} (2007) 3583
  [arXiv:hep-th/0609157].
  %%CITATION = JPAGB,A40,3583;%%
  
%\cite{Fiamberti:2007rj}
\bibitem{w1}
  F.~Fiamberti, A.~Santambrogio, C.~Sieg and D.~Zanon,
  ``Wrapping at four loops in N=4 SYM,''
  arXiv:0712.3522 [hep-th].
  %%CITATION = ARXIV:0712.3522;%%

  %\cite{Keeler:2008ce}
\bibitem{w2}
  C.~A.~Keeler and N.~Mann,
  ``Wrapping Interactions and the Konishi Operator,''
  arXiv:0801.1661 [hep-th].
  %%CITATION = ARXIV:0801.1661;%%
  
  %\cite{Eden:2007rd}
\bibitem{Eden:2007rd}
  B.~Eden,
  ``Boxing with Konishi,''
  arXiv:0712.3513 [hep-th].
  %%CITATION = ARXIV:0712.3513;%%
  
  %\cite{Kotikov:2007cy}
\bibitem{Kotikov:2007cy}
  A.~V.~Kotikov, L.~N.~Lipatov, A.~Rej, M.~Staudacher and V.~N.~Velizhanin,
  ``Dressing and Wrapping,''
  J.\ Stat.\ Mech.\  {\bf 0710}, P10003 (2007)
  [arXiv:0704.3586 [hep-th]].
  %%CITATION = JSTAT,0710,P10003;%%
  
  %\cite{Rej:2005qt}
\bibitem{Hubbard}
  A.~Rej, D.~Serban and M.~Staudacher,
  ``Planar N = 4 gauge theory and the Hubbard model,''
  JHEP {\bf 0603}, 018 (2006)
  [arXiv:hep-th/0512077].
  %%CITATION = JHEPA,0603,018;%%
  
  %\cite{Mann:2005ab}
\bibitem{Mann:2005ab}
  N.~Mann and J.~Polchinski,
  ``Bethe ansatz for a quantum supercoset sigma model,''
  Phys.\ Rev.\  D {\bf 72}, 086002 (2005)
  [arXiv:hep-th/0508232].
  %%CITATION = PHRVA,D72,086002;%%
  
  %\cite{Gromov:2006dh}
\bibitem{Gromov:2006dh}
  N.~Gromov, V.~Kazakov, K.~Sakai and P.~Vieira,
  ``Strings as multi-particle states of quantum sigma-models,''
  Nucl.\ Phys.\  B {\bf 764}, 15 (2007)
  [arXiv:hep-th/0603043].
  %%CITATION = NUPHA,B764,15;%%
  
  %\cite{Gromov:2006cq}
\bibitem{Gromov:2006cq}
  N.~Gromov and V.~Kazakov,
  ``Asymptotic Bethe ansatz from string sigma model on S**3 x R,''
  Nucl.\ Phys.\  B {\bf 780}, 143 (2007)
  [arXiv:hep-th/0605026].
  %%CITATION = NUPHA,B780,143;%%
  
  %\cite{Gromov:2007cd}
\bibitem{Gromov:2007cd}
  N.~Gromov and P.~Vieira,
  ``Constructing the AdS/CFT dressing factor,''
  Nucl.\ Phys.\  B {\bf 790}, 72 (2008)
  [arXiv:hep-th/0703266].
  %%CITATION = NUPHA,B790,72;%%
  
  %\cite{Rej:2007vm}
\bibitem{Rej:2007vm}
  A.~Rej, M.~Staudacher and S.~Zieme,
  ``Nesting and dressing,''
  J.\ Stat.\ Mech.\  {\bf 0708}, P08006 (2007)
  [arXiv:hep-th/0702151].
  %%CITATION = JSTAT,0708,P08006;%%
  

  %\cite{Ambjorn:2005wa}
\bibitem{Ambjorn:2005wa}
  J.~Ambjorn, R.~A.~Janik and C.~Kristjansen,
   ``Wrapping interactions and a new source of corrections to the spin-chain  /
  string duality,''
  Nucl.\ Phys.\  B {\bf 736}, 288 (2006)
  [arXiv:hep-th/0510171].
  %%CITATION = NUPHA,B736,288;%%
  
  %\cite{Janik:2007wt}
\bibitem{Janik:2007wt}
  R.~A.~Janik and T.~Lukowski,
  ``Wrapping interactions at strong coupling -- the giant magnon,''
  Phys.\ Rev.\  D {\bf 76}, 126008 (2007)
  [arXiv:0708.2208 [hep-th]].
  %%CITATION = PHRVA,D76,126008;%%

%\cite{Hatsuda:2008gd,Gromov:2008ie,Heller:2008at}
\bibitem{Hatsuda:2008gd}
  Y.~Hatsuda and R.~Suzuki,
  ``Finite-Size Effects for Dyonic Giant Magnons,''
  arXiv:0801.0747 [hep-th].
  %%CITATION = ARXIV:0801.0747;%%
  
  %\cite{Gromov:2008ie}
\bibitem{Gromov:2008ie}
  N.~Gromov, S.~Schafer-Nameki and P.~Vieira,
  ``Quantum Wrapped Giant Magnon,''
  arXiv:0801.3671 [hep-th].
  %%CITATION = ARXIV:0801.3671;%%
  
  %\cite{Heller:2008at}
\bibitem{Heller:2008at}
  M.~P.~Heller, R.~A.~Janik and T.~Lukowski,
  ``A new derivation of Luscher F-term and fluctuations around the giant
  magnon,''
  arXiv:0801.4463 [hep-th].
  %%CITATION = ARXIV:0801.4463;%%
  
   %\cite{Luscher:1985dn,Klassen:1990ub}
\bibitem{Luscher:1985dn}
  M.~Luscher,
   ``Volume Dependence Of The Energy Spectrum In Massive Quantum Field Theories.
  1. Stable Particle States,''
  Commun.\ Math.\ Phys.\  {\bf 104}, 177 (1986).
  %%CITATION = CMPHA,104,177;%%

%\cite{Klassen:1990ub}
\bibitem{Klassen:1990ub}
  T.~R.~Klassen and E.~Melzer,
   ``On the relation between scattering amplitudes and finite size mass
  corrections in QFT,''
  Nucl.\ Phys.\  B {\bf 362}, 329 (1991).
  %%CITATION = NUPHA,B362,329;%%

%\cite{Arutyunov:2006gs}
\bibitem{Arutyunov:2006gs}
  G.~Arutyunov, S.~Frolov and M.~Zamaklar,
  ``Finite-size effects from giant magnons,''
  Nucl.\ Phys.\  B {\bf 778}, 1 (2007)
  [arXiv:hep-th/0606126].
  %%CITATION = NUPHA,B778,1;%%

%\cite{Astolfi:2007uz}
\bibitem{Astolfi:2007uz}
  D.~Astolfi, V.~Forini, G.~Grignani and G.~W.~Semenoff,
  ``Gauge invariant finite size spectrum of the giant magnon,''
  Phys.\ Lett.\  B {\bf 651}, 329 (2007)
  [arXiv:hep-th/0702043].
  %%CITATION = PHLTA,B651,329;%%

%\cite{Minahan:2008re}
\bibitem{Minahan:2008re}
  J.~A.~Minahan and O.~Ohlsson Sax,
  ``Finite size effects for giant magnons on physical strings,''
  arXiv:0801.2064 [hep-th].
  %%CITATION = ARXIV:0801.2064;%%

%\cite{Hofman:2006xt}
\bibitem{Hofman:2006xt}
  D.~M.~Hofman and J.~M.~Maldacena 
  ``Giant magnons,''
  J.\ Phys.\ A  {\bf 39}, 13095 (2006)
  [arXiv:hep-th/0604135].
  %%CITATION = JPAGB,A39,13095;%%
    
    %\cite{Arutyunov:2007tc}
\bibitem{TBA1}
  G.~Arutyunov and S.~Frolov,
  ``On String S-matrix, Bound States and TBA,''
  JHEP {\bf 0712} (2007) 024
  [arXiv:0710.1568 [hep-th]].
  %%CITATION = JHEPA,0712,024;%%
    
    %\cite{Arutyunov:2008zt}
\bibitem{TBA2}
  G.~Arutyunov and S.~Frolov,
  ``The S-matrix of String Bound States,''
  arXiv:0803.4323 [hep-th].
  %%CITATION = ARXIV:0803.4323;%%
    
  \bibitem{alemaes}
  D.J.~Klein, W.-Seitz, 
  ``Perturbation Expansion of the Linear Hubbard Model,"
  Phys. Rev. B. 8 (1973) 2236. 
  
 \bibitem{LiebWu}
 E.H. Lieb and F.Y. Wu, 
 ``Absence of Mott transition in an exact solution of the  short-range, one-band model in one dimension,"
  Phys. Rev. Lett. 20 (1968) 1445;  Erratum, ibid. 21 (1968) 192; ``The one-dimensional Hubbard   model": A reminiscence," cond-mat/0207529. 
  
  %\cite{Thacker:1974kv}Thacker:1976vp}Klose:2006dd}
\bibitem{Thacker:1974kv}
  H.~B.~Thacker,
  ``Bethe's Hypothesis And Feynman Diagrams: Exact Calculation Of A Three Body
  Scattering Amplitude By Perturbation Theory,''
  Phys.\ Rev.\  D {\bf 11} (1975) 838.
  %%CITATION = PHRVA,D11,838;%%
  
  %\cite{Thacker:1976vp}Klose:2006dd}
\bibitem{Thacker:1976vp}
  H.~B.~Thacker,
  ``Many Body Scattering Processes In A One-Dimensional Boson System,''
  Phys.\ Rev.\  D {\bf 14} (1976) 3508.
  %%CITATION = PHRVA,D14,3508;%%
  
  %\cite{Klose:2006dd}
\bibitem{Klose:2006dd}
  T.~Klose and K.~Zarembo,
  ``Bethe ansatz in stringy sigma models,''
  J.\ Stat.\ Mech.\  {\bf 0605} (2006) P006
  [arXiv:hep-th/0603039].
  %%CITATION = JSTAT,0605,P006;%%
  
%\cite{Dorey:2006dq}
\bibitem{Dorey:2006dq}
  N.~Dorey,
  ``Magnon bound states and the AdS/CFT correspondence,''
  J.\ Phys.\ A  {\bf 39} (2006) 13119
  [arXiv:hep-th/0604175].
  %%CITATION = JPAGB,A39,13119;%%
  
  \bibitem{Bazhanov:1996aq}
  V.~V.~Bazhanov, S.~L.~Lukyanov and A.~B.~Zamolodchikov,
  ``Quantum field theories in finite volume: Excited state energies,''
  Nucl.\ Phys.\  B {\bf 489} (1997) 487
  [arXiv:hep-th/9607099].
  %%CITATION = NUPHA,B489,487;%%
  
 %\cite{SchaferNameki:2006gkmSchaferNameki:2006ey
\bibitem{SchaferNameki:2006gk}
  S.~Schafer-Nameki,
  ``Exact expressions for quantum corrections to spinning strings,''
  Phys.\ Lett.\  B {\bf 639}, 571 (2006)
  [arXiv:hep-th/0602214].
  %%CITATION = PHLTA,B639,571;%%

%\cite{SchaferNameki:2006ey}
\bibitem{SchaferNameki:2006ey}
  S.~Schafer-Nameki, M.~Zamaklar and K.~Zarembo,
  ``How accurate is the quantum string Bethe ansatz?,''
  JHEP {\bf 0612}, 020 (2006)
  [arXiv:hep-th/0610250].
  %%CITATION = JHEPA,0612,020;%%
  
%\cite{Beisert:2004ag,SchaferNameki:2004ik,Beisert:2005di}
\bibitem{Beisert:2004ag}
  N.~Beisert, V.~A.~Kazakov and K.~Sakai,
  ``Algebraic curve for the SO(6) sector of AdS/CFT,''
  Commun.\ Math.\ Phys.\  {\bf 263} (2006) 611
  [arXiv:hep-th/0410253].
  %%CITATION = CMPHA,263,611;%%  

  %\cite{SchaferNameki:2004ik}
\bibitem{SchaferNameki:2004ik}
  S.~Schafer-Nameki,
  ``The algebraic curve of 1-loop planar N = 4 SYM,''
  Nucl.\ Phys.\  B {\bf 714}, 3 (2005)
  [arXiv:hep-th/0412254].
  %%CITATION = NUPHA,B714,3;%%

%\cite{Beisert:2005di}
\bibitem{Beisert:2005di}
  N.~Beisert, V.~A.~Kazakov, K.~Sakai and K.~Zarembo,
  ``Complete spectrum of long operators in N = 4 SYM at one loop,''
  JHEP {\bf 0507}, 030 (2005)
  [arXiv:hep-th/0503200].
  %%CITATION = JHEPA,0507,030;%%

%\cite{Gromov:2007aq}
\bibitem{Gromov:2007aq}
  N.~Gromov and P.~Vieira,
  ``The AdS(5) x S**5 superstring quantum spectrum from the algebraic curve,''
  Nucl.\ Phys.\  B {\bf 789}, 175 (2008)
  [arXiv:hep-th/0703191].
  %%CITATION = NUPHA,B789,175;%%
  %\cite{Gromov:2007aq}
  
%\cite{Gromov:2007ky}
\bibitem{Gromov:2007ky}
  N.~Gromov and P.~Vieira,
 ``Complete 1-loop test of AdS/CFT,''
  JHEP {\bf 0804}, 046 (2008)
  [arXiv:0709.3487 [hep-th]].
  %%CITATION = JHEPA,0804,046;%%

%\cite{Vicedo:2008jy}
\bibitem{Vicedo:2008jy}
  B.~Vicedo,
 ``Semiclassical Quantisation of Finite-Gap Strings,''
  arXiv:0803.1605 [hep-th].
  %%CITATION = ARXIV:0803.1605;%%

\bibitem{GSVfuture}
  N.~Gromov,   S.~Schafer-Nameki, and P.~Vieira,
  To Appear
  %%CITATION = JHEPA,0804,046;%%

\bibitem{Bill}
B. Sutherland, 
 `` A brief history of the quantum soliton with new results on the 
quantization of the Toda lattice,"
Rocky Mtn. J. of Math. 8 (1978) 431. 

%\cite{Beisert:2004ry}
\bibitem{Beisert:2004ry}
  N.~Beisert,
  ``The dilatation operator of N = 4 super Yang-Mills theory and
  integrability,''
  Phys.\ Rept.\  {\bf 405} (2005) 1
  [arXiv:hep-th/0407277].
  %%CITATION = PRPLC,405,1;%%

%\cite{Kulish:1981gi,Faddeev:1996iy}
\bibitem{Kulish:1981gi}
  P.~P.~Kulish, N.~Y.~Reshetikhin and E.~K.~Sklyanin,
  ``Yang-Baxter Equation And Representation Theory. 1,''
  Lett.\ Math.\ Phys.\  {\bf 5} (1981) 393.
  %%CITATION = LMPHD,5,393;%%

  %\cite{Faddeev:1996iy}
\bibitem{Faddeev:1996iy}
  L.~D.~Faddeev,
  ``How Algebraic Bethe Ansatz works for integrable model,''
  arXiv:hep-th/9605187.
  %%CITATION = HEP-TH/9605187;%%
  
  %\cite{Luscher:1982uv}
\bibitem{Luscher:1982uv}
  M.~Luscher,
  ``A New Method To Compute The Spectrum Of Low Lying States In Massless
  Asymptotically Free Field Theories,''
  Phys.\ Lett.\  B {\bf 118} (1982) 391.
  %%CITATION = PHLTA,B118,391;%%
  

\end{thebibliography}

%%%%%%%%%%%%%%%%%%%%%%%%%%%%%%%%%%%%%%%%%%%%%%%%%%%%%%%
%%%%%%%%%%%%%%%%%%%%%%%%%%%%%%%%%%%%%%%%%%%%%%%%%%%%%%%

\end{document}